\documentclass[10pt,twocolumn,letterpaper]{article}
\usepackage{array}
\usepackage{bbding}
\usepackage[caption=false,font=normalsize,labelfont=sf,textfont=sf]{subfig}
\usepackage{upgreek}
\usepackage{stfloats}
\usepackage[figuresright]{rotating}
\usepackage{verbatim}
\usepackage{multirow}
\usepackage{makecell}
\usepackage{iccv}              % To produce the CAMERA-READY version
\definecolor{iccvblue}{rgb}{0.21,0.49,0.74}
\usepackage[pagebackref,breaklinks,colorlinks,allcolors=iccvblue]{hyperref}

\def\paperID{14279} % *** Enter the Paper ID here
\def\confName{ICCV}
\def\confYear{2025}

\title{G$^{2}$SF: Geometry-Guided Score Fusion for Multimodal Industrial Anomaly Detection}

\author{Chengyu Tao$^1$ \ \   Xuanming Cao$^2$ \ \  Juan Du$^{1,2}$\footnote{Corresponding author}\\
 {\small 1.The Hong Kong University of Science and Technology \ \ 2.The Hong Kong University of Science and Technology (Guangzhou)} \\
{\tt\small ctaoaa@connect.ust.hk, xcao743@connect.hkust-gz.edu.cn, juandu@hkust-gz.edu.cn}
}

\begin{document}
\maketitle
\begin{abstract}
Industrial  quality inspection plays a critical role in modern manufacturing by identifying defective products during production. While single-modality approaches using either 3D point clouds or 2D RGB images suffer from information incompleteness, multimodal anomaly detection offers promise through the complementary fusion of crossmodal data. However, existing methods face challenges in effectively integrating unimodal results and improving discriminative power. To address these limitations, we first reinterpret memory bank-based anomaly scores in single modalities as isotropic Euclidean distances in local feature spaces. Dynamically evolving from Euclidean metrics, we propose a novel \underline{G}eometry-\underline{G}uided \underline{S}core \underline{F}usion (G$^{2}$SF) framework that progressively learns an anisotropic local distance metric as a unified score for the fusion task. Through a geometric encoding operator, a novel Local Scale Prediction Network (LSPN) is proposed to predict direction-aware scaling factors that characterize first-order local feature distributions, thereby enhancing discrimination between normal and anomalous patterns. Additionally, we develop specialized loss functions and score aggregation strategy from geometric priors to ensure both metric generalization and efficacy. Comprehensive evaluations on the MVTec-3D AD and Eyecandies datasets demonstrate the state-of-the-art detection performance of our method, and detailed ablation analysis validates each component's contribution. Our code is available at \hyperref[https://github.com/ctaoaa/G2SF]{https://github.com/\allowdisplaybreaks ctaoaa/G2SF.} \footnote{$^*$Corresponding author}
\end{abstract}    
\section{Introduction}
\label{sec:introcution}

Surface quality inspection is crucial in modern industries, as defects like scratches, dents, or cracks can severely impact product functionality and longevity, requiring automated and intelligent inspection systems for large-scale production.  Since defective products are always rare in real-world scenarios, it has led to growing interest in unsupervised anomaly detection \cite{bergmann2019mvtec,bergmann2021mvtec}, which uses only nominal (anomaly-free) samples for training, offering a practical solution for a wide range of industrial applications.

\begin{figure}[!t]
\centering 
\includegraphics[width=1.0\linewidth]{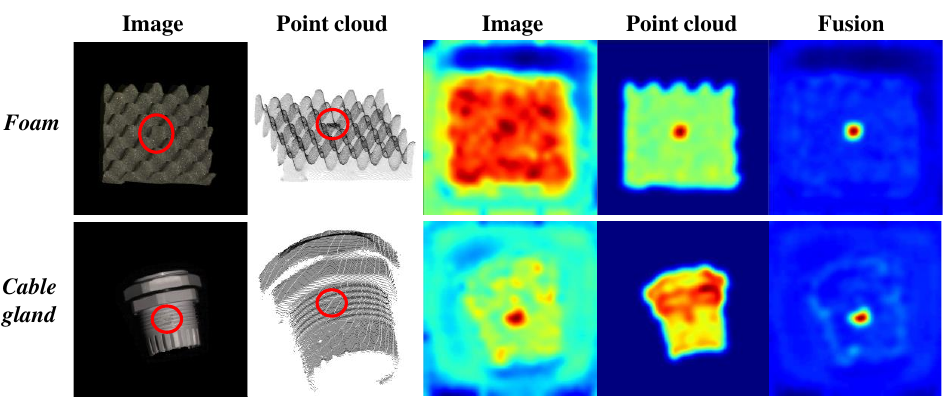}
\caption{Two representative examples from the MVTec-3D AD dataset \cite{bergmann2021mvtec}, including the original RGB image, 3D point cloud, and anomaly score maps from single modalities (image and point cloud) and multimodal fusion (the proposed G$^{2}$SF). \textbf{Top}: Foam with a `cut' anomaly, which is nearly invisible in the RGB image due to its complex texture. \textbf{Bottom}: Cable gland with a `thread' anomaly, which is insufficiently discriminative in the point cloud as it only slightly alters the surface geometry. }
\label{fig:intro_multimodal}
\end{figure}

In complex industrial scenarios, unimodal anomaly detection methods often fail to comprehensively characterize defects due to the limitations of single data sources \cite{horwitz2023back}. For instance, while 3D point clouds provide rich geometric details, they lack texture and color information in RGB images. By integrating these complementary modalities, multimodal approaches achieve superior performance.
Fig. \ref{fig:intro_multimodal} illustrates two representative extreme scenarios from the MVTec-3D AD dataset \cite{bergmann2021mvtec}, where anomalies are predominantly undetectable in one specific modality.

Feature-based anomaly detection approaches have gained popularity due to their flexibility and strong performance \cite{guo2023mldfr,roth2022towards,tu2024self,horwitz2023back,wang2023multimodal}. These methods model the distributions of extracted features, obtained by either handcrafted descriptors or frozen networks, and assign feature-wise anomaly scores. For example, the memory bank-based methods, e.g., PatchCore \cite{roth2022towards}, construct a memory bank of representative prototypes from the whole training dataset and define the anomaly score as the distance to the nearest prototype in the memory bank.

Despite efforts in multimodal methods to learn intermodal representations or mining crossmodal relationships \cite{Asad2M3DF, costanzino2024multimodal, wang2023multimodal}, leveraging the valuable unimodal detection results complementary to the above intermodal information remains underexploited. Current strategies—such as score addition \cite{horwitz2023back}, max selection \cite{Chu2023Shape}, or One-Class Support Vector Machine (OCSVM) \cite{wang2023multimodal}—face a critical limitation: unimodal anomaly scores frequently demonstrate insufficient discriminative capacity in practical applications. As visualized by Fig. \ref{fig:intro_multimodal}, even normal regions in both point cloud and image data exhibit unexpectedly elevated anomaly scores. These unreliable unimodal assessments consequently compromise the fusion performance.
To obtain more discriminative anomaly scores, feature adaptation \cite{reiss2021panda, tu2024self} can be applied that adapts original features into more compact representations beforehand, followed by standard methods, e.g., PatchCore \cite{roth2022towards}, to calculate final anomaly scores. One limitation is that critical details of original features essential for anomaly detection may be over-confidently compromised during first-step adaptation. 

\begin{figure}[!t]
\centering 
\includegraphics[width=\linewidth]{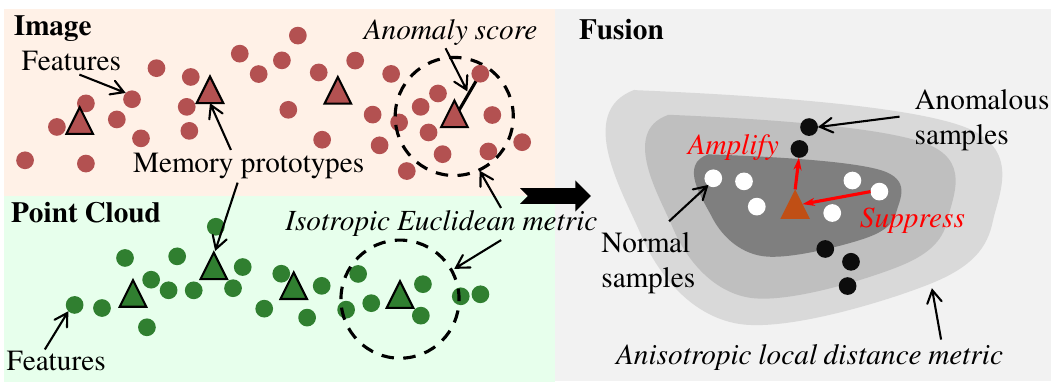}
\caption{\textbf{Left}: The anomaly score in memory bank-based methods, which is defined by the isotropic Euclidean distance to the nearest memory prototype. \textbf{Right}: The anisotropic local distance metric learned by G$^{2}$SF, suppressing anomaly scores for normal samples while amplifying the discrepancy for anomalies.}
\label{fig:motivation}
\end{figure}

This paper re-examines memory bank-based unimodal approaches, revealing a fundamental geometric limitation for their insufficient discriminative capacity: Existing methods compute anomaly scores  via isotropic Euclidean distances in local feature spaces centered on memory prototypes, as illustrated in Fig. \ref{fig:motivation}, but critically ignore other local geometric characteristics ( e.g., directional patterns) essential for anomaly detection.
To address this limitation, we propose Geometry-Guided Score Fusion (G$^{2}$SF) - a framework that replaces isotropic distances with a learned anisotropic local distance metric for multimodal anomaly detection. Specifically, we introduce a Local Scale Prediction Network (LSPN) that derives direction-aware scaling factors to characterize local directional distributions of both modalities in the joint feature space. Subsequently, we formulate a fused metric combining these scales with original Euclidean distances from two modalities to amplify normal/anomaly discrepancies, as shown in Fig. \ref{fig:motivation}. 

Despite the enhanced discriminative power, our G$^{2}$SF offers two other advantages: 
(\romannumeral1) Instead of learning a new metric or anomaly detector from scratch, 
which are often risky in high-dimensional feature spaces \cite{fan2012road,fan2011high}, perhaps increasing with multiple modalities involved, G$^{2}$SF 
explicitly quantifies how unimodal anomaly scores and local geometric structures in feature spaces contribute to the final metric. Consequently, it enables a progressive evolution from the Euclidean metric to an anisotropic metric, mitigating overfitting and ensuring robust metric learning (see Section \ref{subsec:lspn}). (\romannumeral2) As a one-stage anomaly detector, G$^{2}$SF maintains complete modality-specific information throughout the detection process (details are discussed in Section \ref{subsec:encoding}), in contrast to existing feature adaptation frameworks \cite{reiss2021panda, tu2024self} aforementioned above.

The contributions of this paper are summarized as follows:
\begin{itemize}

    \item We propose a systematic G$^{2}$SF framework for industrial multimodal anomaly detection by learning a unified discriminative metric in high-dimensional feature space.

    \item We propose a novel metric deformation paradigm, evolving from isotropic Euclidean metric to an anisotropic one. Specifically, throughout geometric encoding, we design a new LSPN to predict direction-aware scaling factors, subsequently formulating the final metric.
    
    \item We propose critical loss functions to maintain specific geometric properties (e.g., metric consistency and crossmodal correspondence) during metric learning and a novel geometry-informed anomaly scoring strategy, enabling precise multimodal anomaly detection.

    % \item Our method achieves state-of-the-art pixel-wise anomaly segmentation performance (97.9\% AUPRO@30\% \& 99.7\%  P-AUROC) while simultaneously attaining high image-level detection accuracy (97.1\% I-AUROC) on MVTec-3D AD dataset \cite{bergmann2021mvtec}.
    \item Our method achieves state-of-the-art performance on the MVTec 3D-AD and Eyecandies datasets, with particularly notable superiority in the AUPRO@1\% metric.

\end{itemize}
\section{Literature Review}
\label{sec:review}

\subsection{2D Image Anomaly Detection}
\label{sec:review_image}

Unsupervised anomaly detection methodologies can be broadly categorized into two paradigms: \textit{reconstruction}-based approaches and \textit{feature}-based techniques. Reconstruction-based methods typically employ generative architectures, e.g., transformers \cite{pirnay2022inpainting} and diffusion models \cite{wyatt2022anoddpm}, to establish pixel-level reconstruction fidelity for normal images. For inference, anomaly detection is performed by calculating per-pixel discrepancies between input images and their reconstructed counterparts. 

Feature-based approaches, in contrast, circumvent pixel-level comparisons by operating on informative handcraft or deep features. The central challenge in this paradigm involves effectively modeling the distribution of normal features. 
Existing solutions adopt different modeling strategies: PaDiM \cite{defard2021padim} represents features as multivariate Gaussian distributions across spatial positions, while CFlow-AD \cite{gudovskiy2022cflow} employs conditional normalizing flows to capture more complex probability distributions. PNI \cite{bae2023pni} further enhances modeling by incorporating positional and neighborhood relationships of features. 
A notable method is PatchCore \cite{roth2022towards}, which constructs a memory bank of representative normal features through coreset sampling. Owing to its robustness and adaptability, PatchCore has become a foundational method applicable to both 3D \cite{horwitz2023back}, multimodal \cite{horwitz2023back,wang2023multimodal,tu2024self}, and noisy scenarios \cite{jiang2022softpatch}.

\subsection{Multimodal 2D/3D Anomaly Detection}
\label{sec:review_multimodal}

 The pioneering MVTec-3D AD dataset \cite{bergmann2021mvtec} benchmarks this task with various generative networks.  For real-time requirements, Easy-Net \cite{chen2023easynet} designs a multi-scale multimodal autoencoder.
 Feature-based methods prevail in this domain. CFM \cite{costanzino2024multimodal} establishes explicit one-to-one feature mappings to model crossmodal relationships, while 2M3DF \cite{Asad2M3DF} extends this by learning intermodal features via multi-view RGB image generation. Similarly, M3DM \cite{wang2023multimodal} incorporates mutual information for feature fusion. On the other hand, unimodal anomaly scores-containing essential modality-specific information-also demand dedicated fusion strategies. BTF \cite{horwitz2023back} concatenates 2D/3D features with PatchCore \cite{roth2022towards}, equivalent to linear score fusion. Based on the signed distance function (SDF) for 3D patch representation,
Shape-Guided \cite{Chu2023Shape} establishes SDF-guided memory banks for fine-grained detection on image modality but relies on simplistic maximum value between image and SDF scores. M3DM \cite{wang2023multimodal} employs one-class SVM for data-driven score fusion of the original and fused features. However, the inherent limitations in the discriminative power of unimodal anomaly scores remain unaddressed. Toward this deficiency, LSFA \cite{tu2024self} proposes a crossmodal feature adaptation paradigm that learns task-oriented features but risks overlooking the critical details of original features through adaptation.  Finally, at data-level, DAUP \cite{Li2024DAUP} attributes the issue of indiscriminative point cloud scores to the irregular point density and tackles it by resampling point clouds. 

Our G$^{2}$SF introduces two core advances over these existing approaches:
(\romannumeral1) We learn a unified discriminative metric, eliminating suboptimal aggregation from unreliable unimodal anomaly scores;
(\romannumeral2) We introduce a dynamic metric deformation paradigm through isotropic-to-anisotropic evolution, preserving original modality-specific details while enabling precise discrimination in high-dimensional space.
\section{Methodology}
\label{sec:method}

\begin{figure*}[!t]
\centering 
\includegraphics[width=6.8in]{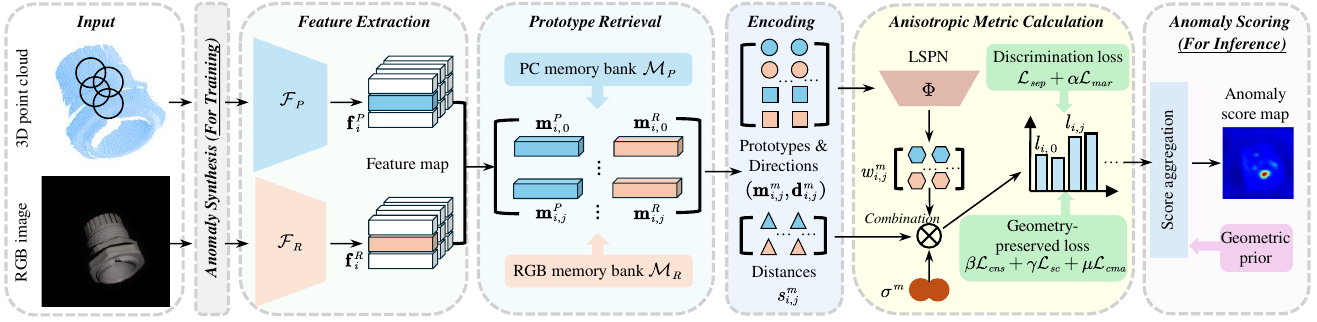}
\caption{Overall pipeline of the proposed G$^{2}$SF. During training, our model takes a pair of 3D point clouds and RGB images as inputs. The anomaly synthesis module (Section \ref{subsec:anomaly_syn}) then injects synthetic anomalies into both modalities. Next, the nearest memory prototypes $\{\mathbf{m}^{P}_{i,j},\mathbf{m}^{R}_{i,j}\}$  are retrieved from the 3D/2D memory banks$\mathcal{M}_P$/$\mathcal{M}_R$ (Section \ref{subsubsec:memory_bank}). In Section \ref{subsec:encoding}, the features
$\mathbf{f}^{P}_{i}/\mathbf{f}^{R}_{i}$ are transformed into geometrically meaningful encodings, represented as triplets: $\langle$ prototypes $\mathbf{m}^{m}_{i,j}$, directions $\mathbf{d}^{m}_{i,j}$, distances $s^{m}_{i,j}$ $\rangle$ ($m \in \{P,R\}$). The LSPN $\Phi$ learns direction-aware scaling factors $w^{m}_{i,j}$ from $(\mathbf{m}^{P}_{i,j},\mathbf{d}^{P}_{i,j},\mathbf{m}^{R}_{i,j},\mathbf{d}^{R}_{i,j})$, which are combined with the distances  $s^{m}_{i,j}$ and global scaling factors $\sigma^{m}$ to compute the anisotropic local distance metric values $\{l_{i,j}\}$, as detailed in Section \ref{subsec:lspn}.
The learning process is supervised by two loss functions: the discrimination loss ($\mathcal{L}_{sep} + \alpha \mathcal{L}_{mar}$, Section \ref{subsubsec:discr_loss}) and the geometry-preserved loss ($\beta \mathcal{L}_{cns} + \gamma \mathcal{L}_{sc} + \mu\mathcal{L}_{cma}$, Section \ref{subsubsec:geo_loss}). Finally, during inference, a score aggregation strategy is applied, leveraging geometric priors for anomaly scoring, as described in Section \ref{subsec:anomaly_scoring}.} 
\label{fig:pipeline}
\end{figure*}

Our G$^{2}$SF aims to learn a unified anisotropic local distance metric for multimodal anomaly detection. The overall pipeline is illustrated in Fig. \ref{fig:pipeline}.

\subsection{Feature Extraction}
\label{subsubsec:feature_extraction}
Our framework accepts a pair of 3D point cloud and RGB image as input. Specifically, we employ two frozen neural networks ($\mathcal{F}_{P}$ and $\mathcal{F}_{R}$) to extract corresponding features $\mathbf{f}^{P}_{i}$ (point cloud) and $\mathbf{f}^{R}_{i}$ (image) at location $i$.
Following existing methods \cite{wang2023multimodal,costanzino2024multimodal,tu2024self}, we adopt two vision transformers, DINO \cite{caron2021emerging} for images and Point-MAE \cite{pang2022masked} for point clouds, considering their powerful representational capacity.  $\mathbf{f}^{P}_{i}$ and $\mathbf{f}^{R}_{i}$ have high dimensions, i.e., 1152 and 768, respectively.

\subsection{Prototype Retrieval from Memory Bank}
\label{subsubsec:memory_bank}

Our G$^{2}$SF builds on the memory bank-based methods for unimodal anomaly detection. 
For each modality $m\in{P,R}$, we construct a memory bank $\mathcal{M}^m$ that collects prototypical features from the entire training dataset following PatchCore \cite{roth2022towards}. Subsequently, for a feature $\mathbf{f}^{m}_{i}$, we can retrieve its $2k+1$ nearest neighbors  $\{\mathbf{m}^{m}_{i,j}\}_{j=0}^{2k}$ from $\mathcal{M}^m$. Critically, the Euclidean distance between $\mathbf{f}^{m}_{i}$ and the nearest $\mathbf{m}^{m}_{i,0}$ is typically defined as the anomaly score $s^{m}_{i,0}$ in these memory bank-based methods \cite{wang2023multimodal}.

\subsection{Geometric Feature Encoding}
\label{subsec:encoding}
The main issue that limits the discriminative capability of $s^{m}_{i,0}$ in Section \ref{subsubsec:memory_bank} is its isotropic nature, which ignores other critical characteristics valuable for anomaly detection. Nevertheless, local directional consistency can act as an auxiliary property to quantitatively evaluate the normality of target samples: Features aligned with the principal directions of normal samples within the local feature spaces are expected to demonstrate a higher degree of normality. 

Therefore, a geometric encoding operator $\mathcal{E}(\cdot)$ that naturally characterizes the first-order local distribution within the local manifold centered at $\mathbf{m}^{m}_{i,j}$ can be defined as \cite{zhou2021feature}:
\begin{equation}
\begin{split}
    (\mathbf{m}^{m}_{i,j}, \mathbf{d}^{m}_{i,j}, s^{m}_{i,j}) &=\mathcal{E}_{{\mathbf{m}^{m}_{i,j}}}(\mathbf{f}^{m}_{i}), \\
    s^{m}_{i,j} = \|\mathbf{f}^{m}_{i}-\mathbf{m}^{m}_{i,j}\|, \ \ &\mathbf{d}^{m}_{i,j} = (\mathbf{f}^{m}_{i}-\mathbf{m}^{m}_{i,j})/s^{m}_{i,j},
\end{split}
\label{eq:encoding}
\end{equation}
where $s^{m}_{i,j}$ represents the Euclidean distance and $\mathbf{d}^{m}_{i,j}$ denotes the directional vector. Moreover, Eq. \eqref{eq:encoding} is a seamless encoding of $\mathbf{f}^{m}_{i}$. This property enables the preservation of complete details for the subsequent learning process, avoiding comprising valuable information in feature adaptation frameworks \cite{reiss2021panda,tu2024self}.

\subsection{Anisotropic Local Distance Metric}
\label{subsec:lspn}

Based on Eq. \eqref{eq:encoding}, we aim to learn a unified metric $l(\mathbf{f}^{P}_{i}, \mathbf{f}^{R}_{i}, \mathbf{m}^{P}_{i,j}, \mathbf{m}^{R}_{i,j})$ for multimodal anomaly detection that outperforms the Euclidean distances $s^{P}_{i,j}$ and  $s^{R}_{i,j}$.  While existing local metric learning methods \cite{wang2012parametric,ye2019learning} use instance-dependent Mahalanobis distances, they are computationally expensive and unsuitable for our high-dimensional joint feature space (1920 dimensions). From the perspective of spectral analysis, Mahalanobis distances determine the scaling factors along principle directions. Mimicking this desired property, we propose the LSPN that leverages a neural network for computational efficiency and implicitly captures shared information across local spaces.

Specifically, LSPN predicts direction-aware scaling factors $w^{m}_{i,j}$ via a simple Multilayer Perceptron $\Phi(\cdot)$ (MLP) as:
\begin{equation}
    [w^{P}_{i,j},w^{R}_{i,j}] = \Phi(\mathbf{m}^{P}_{i,j}, \mathbf{m}^{R}_{i,j}, \mathbf{d}^{P}_{i,j}, \mathbf{d}^{R}_{i,j}).
\label{eq:weight}
\end{equation}
Furthermore, the anisotropic metric $l(\cdot)$ is defined in terms of these scaling factors as:
\begin{equation}
   l_{i,j} = l(\mathbf{f}^{P}_{i}, \mathbf{f}^{R}_{i}, \mathbf{m}^{P}_{i,j}, \mathbf{m}^{R}_{i,j}) = \sum_{m\in \{P,R\}} w^{m}_{i,j}s^{m}_{i,j}\sigma^m,
\label{eq:metric}
\end{equation}
where $\sigma^m$ is a \textit{trainable global} scaling factor that describes the influence of modality $m$.  Eq. \eqref{eq:metric} demonstrates that $l(\cdot)$ incorporates structural priors from unimodal anomaly detection, where $s^{m}_{i,0}$ explicitly represents the unimodal anomaly score. When initializing 
$\Phi$ with $w^{m}_{i,j} \approx 1$, the metric $l(\cdot)$ originates as a Euclidean measure in the joint feature space. Although this baseline configuration is marginally effective in unimodal scenarios, it progressively evolves into an anisotropic metric during training under the regularization of $s^{m}_{i,j}$. This unique property prevents overfitting that would occur when learning a metric entirely from scratch \cite{fan2011high,fan2012road}. 
Furthermore, the supplementary material tracks model performance throughout different training epochs, revealing that G$^{2}$SF attains competitive performance within remarkably few training epochs that benefit from the above design.

\begin{figure}[!t]
\centering 
\includegraphics[width=\linewidth]{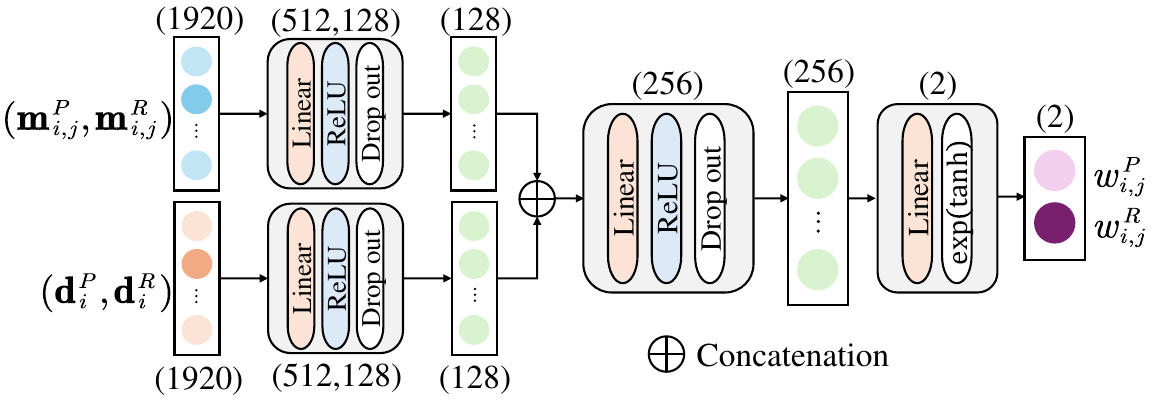}
\caption{The architecture of LSPN.
Each block contains a linear layer, ReLU activation, and dropout. The final layer activation is a $\exp({\tanh{(\cdot)}})$ function that is symmetrical to one and serves as scale suppression/amplification.}
\label{fig:lspn_design}
\end{figure}

The architecture of $\Phi(\cdot)$ is illustrated in Fig. \ref{fig:lspn_design}, which consists of two parallel branches that independently concatenate and encode the prototypes $(\mathbf{m}^{P}_{i,j},\mathbf{m}^{R}_{i,j})$ and directions $(\mathbf{d}^{P}_{i,j},\mathbf{d}^{R}_{i,j})$. $w^{m}_{i,j}$ is obtained
through an activation $\exp({\tanh{(\cdot)}})$ at the final layer, ensuring $w^{m}_{i,j} \in [e^{-1}, e^1]$. This constrains the scaling factors to be symmetric around the Euclidean baseline ($w^{m}_{i,j} = 1$). To enhance the discriminative power of $l(\cdot)$, we aim to reduce scales ($w^{m}_{i,j} < 1$) for normal samples, while increasing scales ($w^{m}_{i,j} > 1$) for anomalous samples.

\subsection{Loss Functions}
\label{subsec:loss}
To ensure stable training,  we employ synthetic anomaly injection in Section \ref{subsec:anomaly_syn}. We denote the label $y_i \in \{0,1\}$ for $(\mathbf{f}^{P}_{i},\mathbf{f}^{R}_{i})$, where $y_i=1$ denotes synthetic anomalies. Additionally,
we denote $(\cdot)_+$  as $\max(\cdot, 0)$. We deploy the discrimination losses ($\mathcal{L}_{sep},\mathcal{L}_{mar}$) in Section \ref{subsubsec:discr_loss} to ensure the learned $l(\cdot)$ capable of discriminating normal ($y_i=0$) and anomalous ($y_i=1$) samples. Furthermore, the geometry-preserved losses ($\mathcal{L}_{cns},\mathcal{L}_{sc},\mathcal{L}_{cma}$) are proposed in Section \ref{subsubsec:geo_loss} to ensure effective metric learning by incorporating desired geometric properties related to metric consistency, local scaling factors, and crossmodal alignments.

\subsubsection{Discrimination Loss}
\label{subsubsec:discr_loss}

We prioritize $l_{i,0}$ to inherit its discriminative capability w.r.t. the nearest prototype $\boldsymbol{m}^{m}_{i,0}$.
 
\noindent\textbf{Separation Loss}: 
The basic separation loss $\mathcal{L}_{sep}$ is to minimize $l_{i,0}$ for normal samples while simultaneously maximizing them for anomalous instances, following the semi-supervised anomaly detection approach \cite{ruff2020deep}:
\begin{equation}
\mathcal{L}_{sep} = \sum_i (1-y_i)l_{i,0} + y_i(l^{-1}_{i,0} - m_0^{-1})_+,
\label{eq:sep_loss}
\end{equation}
where $m_0$ prevents excessively large anomaly scores.

\noindent\textbf{Margin Loss}: The distribution overlap between normal and anomalous samples is also crucial for anomaly detection \cite{cao2023collaborative}, which is considered by the margin loss $\mathcal{L}_{mar}$ as:
{
\small
\begin{equation}
    \mathcal{L}_{mar} = \frac{1}{m_u} \sum_i  \underbrace{(1-y_i)(l_{i,0}-m_l)_+}_{\text{Overlap of Normal Part}}
    + \underbrace{y_i(m_u-l_{i,0})_+}_{\text{Overlap of Anomalous Part}},
\label{eq:margin_loss}
\end{equation}}where $m_l = \min_{i|y_i=1} l_{i,0}$ is the lower bound of $l_{i,0}$ for anomalous samples and $m_u= \max_{i|y_i=0} l_{i,0}$ is the upper bound for normal samples.

\subsubsection{Geometry-Preserved Loss}
\label{subsubsec:geo_loss}

Since learning a robust and generalizable $l(\cdot)$ in the high-dimensional space for anomaly detection is challenging, despite the architecture design in Section \ref{subsec:lspn}, we aim to exploit the desired geometric properties of $l(\cdot)$ that benefit our task.

\noindent\textbf{Consistency Loss}: 
Since Eqs. \eqref{eq:sep_loss} and \eqref{eq:margin_loss} focus on the discriminative power of $l_{i,0}$, we exploit the geometric consistency of $l_{i,j}$ across different local spaces around $\mathbf{m}^{m}_{i,j}$, including the following two properties:
\begin{itemize}
    \item \textit{Proximal consistency}: $l(\cdot)$ is continuous across neighborhood. For adjacent local spaces, i.e., $1\leq j\leq k$, $l_{i,j}$ is upper bounded by $\eta_{i,j}l_{i,0}$ for $\eta_{i,j} > 1$. It means that the partial gradient of $l(\cdot)$  w.r.t. $(\mathbf{m}^{P}_{i,j},\mathbf{m}^{R}_{i,j})$ is limited.
    
    \item \textit{Distal separation}: For sufficiently large Euclidean distances, $l(\cdot)$ should maintain correspondingly significant values.
    We mandate $l_{i,j}$ in distant local spaces indexed by $k+1\leq j\leq 2k$ lower-bounded by $l_{i,0}$. This constraint prevents metric collapse in high-dimensional spaces to avoid incredible values.
\end{itemize} 
Accordingly, we formulate the consistency loss $\mathcal{L}_{cns}$ as:
{\small
\begin{equation}
\begin{split}
    \mathcal{L}_{cns} = {\frac{1}{k}}\sum_i  \underbrace{\sum_{1\leq j\leq k} (l_{i,j} -\eta_{i,j} l_{i,0})_+}_{\text{Proximal Consistency}} + 
    \underbrace{\sum_{k+1\leq j\leq2k} (l_{i,0} -l_{i,j})_+}_{\text{Distal Separation}}.
    \label{eq:cns_loss}
\end{split}
\end{equation}
}Despite the spatial regularization, $\mathcal{L}_{cns}$ also effectively increases sample diversity within each local space through multi-prototype observations as data augmentation.

\noindent\textbf{Scaling Loss}:
Despite $l(\cdot)$, we aim to encourage $w^{m}_{i,0}$ to encode local directional distributions explicitly, that is, suppress/amplify for normal/anomalous samples. To achieve this,  we propose an asymmetric scaling loss:
\begin{equation}
    \mathcal{L}_{sc} = \sum_i\sum_{m\in\{P,R\}} (1-y_i) (w^{m}_{i,0}-1)_{+} + y_i(e^{1}-w^{m}_{i,0}).
\label{eq:sc_loss}
\end{equation}
$\mathcal{L}_{sc}$ asymmetrically compresses scaling factors of normal samples while confining those of synthetic anomalies near the upper limit $e^{1}$.  This design addresses the more pronounced deviations of synthetic anomalies than real-world ones, imposing more penalties on them.

\noindent\textbf{Crossmodal Alignment Loss}: 
Currently, both prototypes and directions are naively concatenated as input to LSPN (Sec. \ref{subsec:lspn}), ignoring their crossmodal correspondence. To resolve this, we propose a simple 
strategy by disentangling this property.

Specifically, by permuting indices $\{i\}\rightarrow \{\tilde{i}\}$, we can construct two kinds of negative samples:
\begin{equation}
(\mathbf{m}^{P}_{i,0},\mathbf{m}^{R}_{\tilde{i},0},\mathbf{d}^{P}_{i,0},\mathbf{d}^{R}_{i,0}), \ \ (\mathbf{m}^{P}_{i,0},\mathbf{m}^{R}_{i,0},\mathbf{d}^{P}_{i,0},\mathbf{d}^{R}_{\tilde{i},0}).
\end{equation}
The former breaks holistic prototype coherence. In contrast, the latter focuses on capturing discordant directions while maintaining prototype consistency at a local structural granularity.

During training, we construct the above two kinds with equal probability. Denote $\tilde{w}^{m}_{i,0}$ ($m \in \{P,R\}$) as their associated scaling factors, we define the crossmodal alignment loss as:
\begin{equation}
    \mathcal{L}_{cma} = \sum_i \sum_{m\in\{P,R\}} (e^{1}-\tilde{w}^{m}_{i,0}),
    \label{eq:cma_loss}
\end{equation}
which enforces large scales similar to synthetic anomalous samples in Eq. \eqref{eq:sc_loss}.

\subsubsection{Overall Loss} Finally, the overall loss $\mathcal{L}$ of our G$^{2}$SF framework combines all terms as:
\begin{equation}
    \mathcal{L} = \underbrace{\mathcal{L}_{sep} + \alpha \mathcal{L}_{mar}}_{\text{Discrimination Loss}} + \underbrace{\beta \mathcal{L}_{cns} + \gamma \mathcal{L}_{sc} + \mu  \mathcal{L}_{cma}}_{\text{Geometry-Preserved Loss}},
\end{equation}
where $\alpha$, $\beta$, $\gamma$, and $\mu$ are hyper-parameters balancing the loss components. 

\subsection{Geometry-Informed Anomaly Scoring}
\label{subsec:anomaly_scoring}

During inference, G$^{2}$SF computes the anomaly score $s_i$ for each multimodal feature pair $(\mathbf{f}^{P}_{i},\mathbf{f}^{R}_{i})$. Concretely, by propagating features through the geometric encoding operator and LSPN, we obtain the anisotropic metrics  $\{l_{i,j}\}^{k}_{j=0}$ across 
$k+1$ nearest local spaces, adhering to the proximal consistency constraint in Eq. \eqref{eq:cns_loss}. We apply a $\min$ operator to aggregate $\{l_{i,j}\}^{k}_{j=0}$ to define $s_i$:
\begin{equation}
    s_i = \min \{l_{i,j}|j=0,1,...,k\}.
\label{eq:aggregate}
\end{equation}
This definition stems from a geometric prior. Specifically, memory bank-based methods calculate anomaly scores using Euclidean distances to their memory banks (Section \ref{subsubsec:memory_bank}), whereas our formulation in Eq. \eqref{eq:aggregate} reformulates this as distance measurement to the normal data manifold through the anisotropic metrics $l(\cdot)$. We will compare Eq. \eqref{eq:aggregate} with several alternatives in Section \ref{subsec:ablation}.

Finally, the sample-level anomaly score is obtained by taking the maximum $s_i$ across all feature pairs extracted from a test sample. Following \cite{costanzino2024multimodal,wang2023multimodal,tu2024self}, we apply bilinear interpolation and a Gaussian kernel with size 4 to upsample and smooth the score map, ensuring resolution alignment with input dimensions.

\subsection{Anomaly Synthesis}
\label{subsec:anomaly_syn}

Our anomaly synthesis procedure operates following CutPaste approaches \cite{yao2023explicit,li2021cutpaste,chen2023easynet}: First, we generate Berlin noise maps, which are binarized into mask maps to select paste locations. Based on these masks, we then cut corresponding regions from source (normal) 3D point clouds and RGB images from the training dataset. The source point clouds can be other classes or more general 3D models. Subsequently, we apply random augmentations to corrupt the regions - specifically implementing pixel modifications for source images \cite{yao2023explicit}, e.g., distortion, brightness, and sharpness changes, while executing random translations for source point clouds. The corrupted regions are finally pasted onto target data. Examples of synthetic anomalous samples are provided in the supplementary material.

\begin{table*}[ht]
\centering
\footnotesize

\begin{tabular}{>{\centering\arraybackslash}m{0.2cm} |
    >{\centering\arraybackslash}m{0.6cm} >{\centering\arraybackslash}m{0.6cm} |
    >{\centering\arraybackslash}m{0.6cm} >{\centering\arraybackslash}m{0.6cm}
    >{\centering\arraybackslash}m{0.6cm} >{\centering\arraybackslash}m{0.6cm}
    >{\centering\arraybackslash}m{0.6cm} >{\centering\arraybackslash}m{0.6cm}
    >{\centering\arraybackslash}m{0.6cm} >{\centering\arraybackslash}m{0.6cm}
    >{\centering\arraybackslash}m{0.6cm} >{\centering\arraybackslash}m{0.6cm} |
    >{\centering\arraybackslash}m{0.6cm}} \hline\hline
     & \multicolumn{2}{c|}{\textbf{Method}}  & \textit{Bagel} & \textit{Cable Gland} & \textit{Carrot} & \textit{Cookie} & \textit{Dowel} & \textit{Foam} & \textit{Peach} & \textit{Potato} & \textit{Rope} & \textit{Tire} & \textbf{Mean} \\ \hline
     
   \multirow{10}{*}{\rotatebox{90}{\textbf{I-AUROC}}} & \multicolumn{2}{c|}{DepthGAN\cite{bergmann2021mvtec}} & 0.538 & 0.372 & 0.580 & 0.603 & 0.430 & 0.534 & 0.642 & 0.601 & 0.443 & 0.577 & 0.532 \\
    & \multicolumn{2}{c|}{VoxelGAN\cite{bergmann2021mvtec}} & 0.680 & 0.324 & 0.565 & 0.399 & 0.497 & 0.482 & 0.566 & 0.579 & 0.601 & 0.482 & 0.517 \\
    & \multicolumn{2}{c|}{BTF\cite{horwitz2023back}} & 0.918 & 0.748 & 0.967 & 0.883 & 0.932 & 0.582 & 0.896 & 0.912 & 0.921 & 0.886 & 0.865 \\  
    & \multicolumn{2}{c|}{EasyNet\cite{chen2023easynet}} & 0.991 & \textbf{0.998} & 0.918 & 0.968 & 0.945 & 0.945 & 0.905 & 0.807 & \underline{0.994} & 0.793 & 0.926 \\
 & \multicolumn{2}{c|}{AST\cite{rudolph2023asymmetric}} & 0.983 & 0.873 & 0.976 & 0.971 & 0.932 & 0.885 & 0.974 & 0.981 & \textbf{1.000} & 0.797 & 0.937 \\  
 & \multicolumn{2}{c|}{M3DM \cite{wang2023multimodal}} & 0.994 & 0.909 & 0.972 & 0.976 & 0.960 & 0.942 & 0.973 & 0.899 & 0.972 & 0.850 & 0.945 \\  
 & \multicolumn{2}{c|}{Shape-guided \cite{Chu2023Shape}} & 0.986 & 0.894 & 0.983 & 0.991 & 0.976 & 0.857 & \underline{0.990} & 0.965 & 0.960 & 0.869 & 0.947 \\
 & \multicolumn{2}{c|}{CFM \cite{costanzino2024multimodal}} & 0.994 & 0.888 & 0.984 & \underline{0.993} & \textbf{0.980} & 0.888 & 0.941 & 0.943 & 0.980 & \underline{0.953} & 0.954 \\
 & \multicolumn{2}{c|}{2M3DF \cite{Asad2M3DF}} & 0.992 & \underline{0.969} & \underline{0.988} & 0.985 & 0.981 & 0.947 & 0.979 & 0.942 & 0.976 & 0.898 & 0.966 \\ 
& \multicolumn{2}{c|}{LSFA \cite{tu2024self}} &\textbf{1.000} & 0.939 & 0.982 & 0.989 & 0.961 & \underline{0.951} & 0.983 & 0.962 & 0.989 & \underline{0.951} & \textbf{0.971} \\
& \multicolumn{2}{c|}{DAUP \citep{Li2024DAUP}} & 0.996 &  0.889 &  \textbf{0.996} &  \textbf{0.998} &  \underline{0.977} & 0.939 & 0.983 & \underline{0.986} & 0.979 & \textbf{0.960} & \underline{0.970} \\
& \multicolumn{2}{c|}{Ours} & \underline{0.997} & 0.923 & \underline{0.993} & 0.967& 0.966& \textbf{0.991}& \textbf{0.994}& \textbf{0.988}& 0.966 &0.922 & \textbf{0.971} \\ 

\hline\hline

\multirow{11}{*}{\rotatebox{90}{\textbf{AUPRO@30\%}}} & \multicolumn{2}{c|}{DepthGAN\cite{bergmann2021mvtec}} & 0.421 & 0.422 & 0.778 & 0.696 & 0.494 & 0.252 & 0.285 & 0.362 & 0.402 & 0.631 & 0.474 \\
    & \multicolumn{2}{c|}{VoxelGAN\cite{bergmann2021mvtec}} &0.664 & 0.620 & 0.766 & 0.740 & 0.783 & 0.332 & 0.582 & 0.790 & 0.633 & 0.483 & 0.639 \\
    & \multicolumn{2}{c|}{BTF\cite{horwitz2023back}} &0.976 & 0.969 & 0.979 & 0.973 & 0.933 & 0.888 & 0.975 & 0.981 & 0.950 & 0.971 & 0.959 \\  
 & \multicolumn{2}{c|}{AST\cite{rudolph2023asymmetric}} & 0.970 & 0.947 & 0.981 & 0.939 & 0.913 & 0.906 & 0.979 & \underline{0.982} & 0.889 & 0.940 & 0.944 \\  
 & \multicolumn{2}{c|}{M3DM \cite{wang2023multimodal}} & 0.970 & 0.971 & 0.979 & 0.950 & 0.941 & 0.932 & 0.977 & 0.971 & 0.971 & 0.975 & 0.964 \\  
 & \multicolumn{2}{c|}{Shape-guided \cite{Chu2023Shape}} &0.981& 0.973 &\underline{0.982}& 0.971& 0.962& \textbf{0.978} &0.981 &\textbf{0.983} &0.974 &0.975& \underline{0.976} \\
 & \multicolumn{2}{c|}{CFM \cite{costanzino2024multimodal}} & 0.979 & 0.972 & \underline{0.982} & 0.945 & 0.950 & 0.968 & 0.980 & \underline{0.982} & \underline{0.975} & \underline{0.981} & 0.971 \\
 & \multicolumn{2}{c|}{2M3DF \cite{Asad2M3DF}} & \textbf{0.986} & \textbf{0.983} & \textbf{0.987} & \textbf{0.984} & \underline{0.968} & 0.949 & \textbf{0.986} & 0.971 & \textbf{0.988} & 0.954 & 0.975 \\
& \multicolumn{2}{c|}{LSFA \cite{tu2024self}} &\textbf{0.986}& 0.974& 0.981 & 0.946& 0.925& 0.941& \textbf{0.983}& \textbf{0.983}& 0.974& \textbf{0.983} &0.968 \\
& \multicolumn{2}{c|}{DAUP \citep{Li2024DAUP}} &0.976 & \underline{0.977} &0.980&0.960 &0.924 &0.966 &0.981&0.978 &0.972 &0.980 &0.969\\
& \multicolumn{2}{c|}{Ours} & \underline{0.982} &\underline{0.977} & \underline{0.982} & \underline{0.979} & \textbf{0.971} & \underline{0.976} & 0.982 & \textbf{0.983} & \underline{0.978} & \underline{0.981}& \textbf{0.979} \\ \hline\hline
\end{tabular}
\caption{\textbf{I-AUROC} and \textbf{AUPRO@30\%} results on MVTec-3D AD. Best results in \textbf{bold}, runner-ups \underline{underlined}.}
\label{tab:performance_comparison}
\end{table*}

\section{Experiment}
\label{sec:experiment}

\begin{figure}[!t]
\centering
\includegraphics[width=\linewidth]{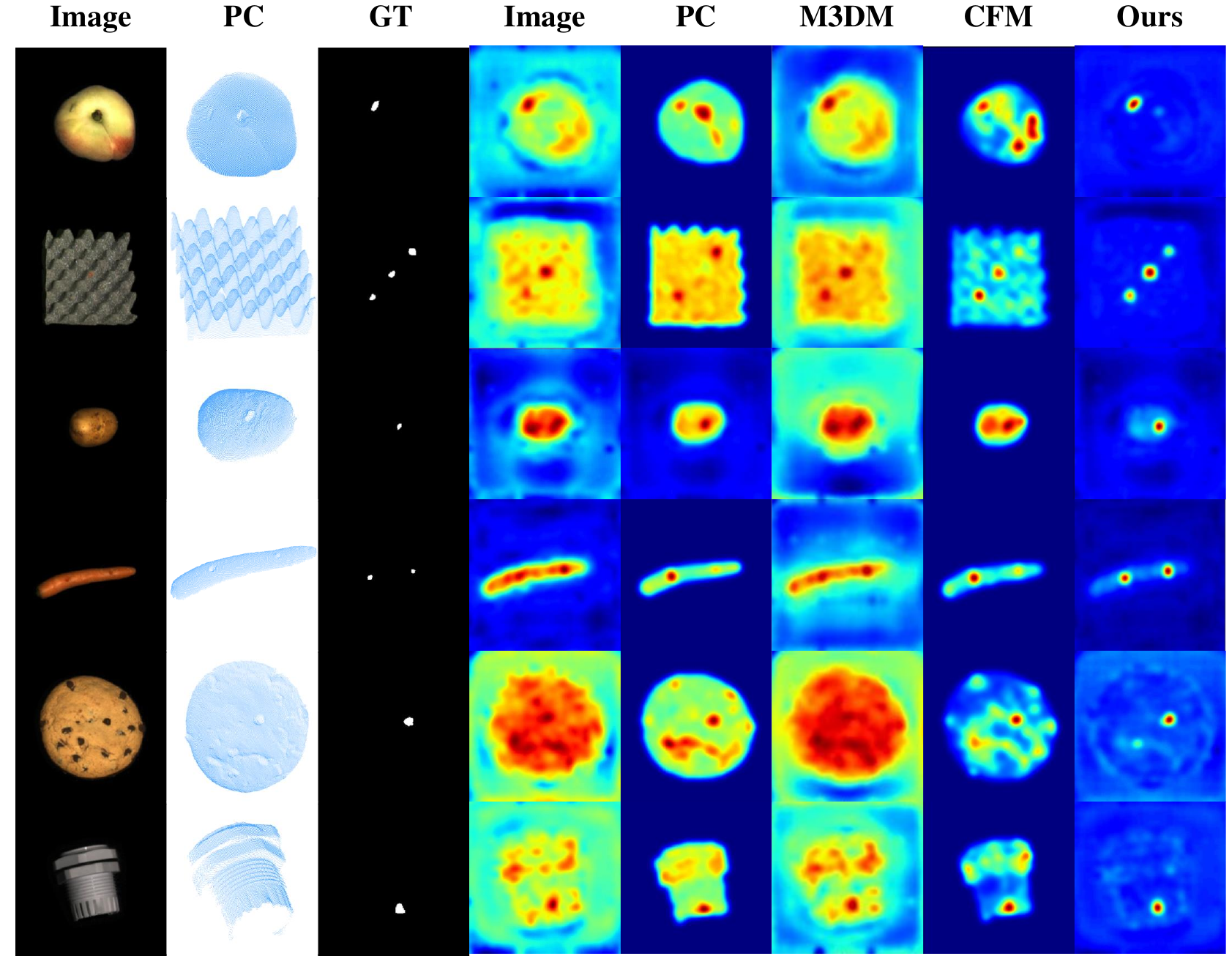}
\caption{Qualitative comparison on MVTec-3D AD  \cite{bergmann2021mvtec}: Visualized results include original RGB images with aligned 3D point clouds (PC) and ground truths (GT), alongside anomaly detection outputs from unimodal methods (PatchCore \cite{roth2022towards} for image/point cloud) and multimodal benchmarks (M3DM \cite{wang2023multimodal}, CFM \cite{costanzino2024multimodal}), compared against our proposed G$^{2}$SF framework.}
\label{fig:qualitative}
\end{figure}

\begin{table}[ht]
\centering
\footnotesize

\begin{tabular}{>{\centering\arraybackslash}m{0.2cm} |
    >{\centering\arraybackslash}m{1.5cm} >{\centering\arraybackslash}m{1.0cm} 
    >{\centering\arraybackslash}m{1.0cm} >{\centering\arraybackslash}m{1.0cm}
    >{\centering\arraybackslash}m{1.0cm}} \hline\hline
    &  {\textbf{Method}} & {I-AUROC} & {P-AUROC} & \makecell{AUPRO\\@30\%} & \makecell{AUPRO\\@1\%}   \\ \hline
   \multirow{8}{*}{\rotatebox{90}{\textbf{MVTec-3D AD} \ \ \ \ }} &BTF \cite{horwitz2023back}  & 0.865 & 0.992 & 0.959 &  0.383 \\ 
   &AST \cite{rudolph2023asymmetric} & 0.937 & 0.976  & 0.944 & 0.398 \\ 
  &M3DM \cite{wang2023multimodal} & 0.945  & 0.992 & 0.964 &  0.394 \\ 
  &Shape. \cite{costanzino2024multimodal} & 0.947  &\underline{0.996} & \underline{0.976} & \underline{0.456} \\ 
  &CFM \cite{costanzino2024multimodal} & 0.954 & 0.993 & 0.971 & 0.455 \\ 
  &2M3DF \cite{Asad2M3DF} & 0.966 & - & 0.975 & - \\ 
  &LSFA \cite{tu2024self} & \textbf{0.971} & 0.993 & 0.968 & - \\ 
   &DUAP \cite{Li2024DAUP} & \underline{0.970} & - & 0.969 & - \\ 
    &Ours & \textbf{0.971} & \textbf{0.997} & \textbf{0.979} & \textbf{0.468} \\ 
\hline\hline
    \multirow{5}{*}{\rotatebox{90}{\textbf{Eyecandies} \ \ \ \ }} & AST \cite{rudolph2023asymmetric}  &0.780  & 0.902 & 0.744  & 0.149 \\
    & M3DM \cite{wang2023multimodal} &0.882 & \underline{0.977} & 0.887 & 0.331  \\
    & CFM \cite{costanzino2024multimodal} & 0.881 & 0.973  &0.887 & \underline{0.335} \\
    & 2M3DF \cite{Asad2M3DF} & \underline{0.897} & - & \underline{0.890} &-  \\
    & LSFA \cite{tu2024self}  & -  &0.974& - & - \\
    &Ours  & \textbf{0.902} & \textbf{0.982} & \textbf{0.898} & \textbf{0.357}  \\ \hline\hline
\end{tabular}
\caption{Mean metrics for all categories on both datasets. }
\label{tab:all_mean}
\end{table}

\subsection{Experimental Details}
\label{subsec:expr_details}

\noindent\textbf{Dataset}: Experiments are conducted on the MVTec-3D AD \cite{bergmann2021mvtec} and Eyecandies\cite{bonfiglioli2022eyecandies} datasets. MVTec-3D AD contains 10 industrial categories (210-361 train, 69-132 test per class), while Eyecandies provides 10 synthetic food items (1k train, 400 test per class). Each sample provides aligned RGB images and 3D point clouds, enabling pixel-wise correspondence between color and spatial coordinates.

\noindent\textbf{Evaluation Metrics}: We assess the image-level anomaly detection performance by the Area Under the Receiver Operator Curve (I-AUROC). For evaluating the pixel-wise anomaly segmentation, we adopt the pixel-level AUROC (P-AUROC) and the
Area Under the Per-Region Overlap (AUPRO) at 30\% integration limits (denoted as AUPRO@30\%), the same as existing methods \cite{costanzino2024multimodal,horwitz2023back,Chu2023Shape,wang2023multimodal}. 
To better demonstrate the advantages of our method in practical industrial scenarios with lower false positive rates, we also adopted the AUPRO@1\% metric (additional AUPRO@5\% and AUPRO10\% in the supplementary material) used in recent studies \cite{Chu2023Shape,costanzino2024multimodal}.

\noindent\textbf{Implementation}: We follow a similar procedure of M3DM \cite{wang2023multimodal} for feature extraction and preprocessing of the dataset.  For $\mathcal{L}_{cns}$, we set $k=5$ and $\eta_{0}=1.2$. We train our G$^{2}$SF with the Adam optimizer \cite{kingma2014adam} with 80 epochs on Linux Nvidia RTX 4090. More details are provided in the supplementary material.

\begin{figure}[!t]
\centering
\includegraphics[width=\linewidth]{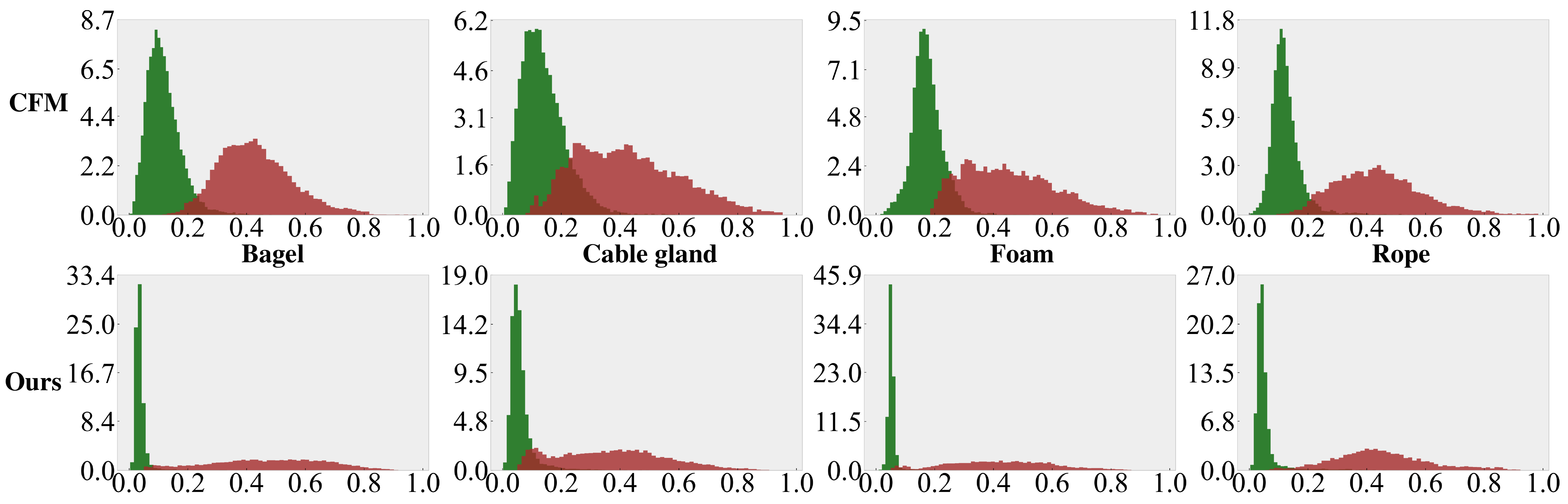}
\caption{Pixel-wise anomaly score distributions in \textit{Bagel}, \textit{Cable gland}, \textit{Foam}, and \textit{Rope} categories, where the horizontal axis is normalized score (0-1) with vertical axis as the probabilistic density.
Green/Brown distributions represent normal/anomalous regions respectively. Compared with CFM \cite{costanzino2024multimodal} (top), our method (bottom) achieves significant distribution compression in normal regions while maintaining anomaly discrimination.}
\label{fig:distribution}
\end{figure}

\begin{figure}[!t]
\centering
\includegraphics[width=1.0\linewidth]{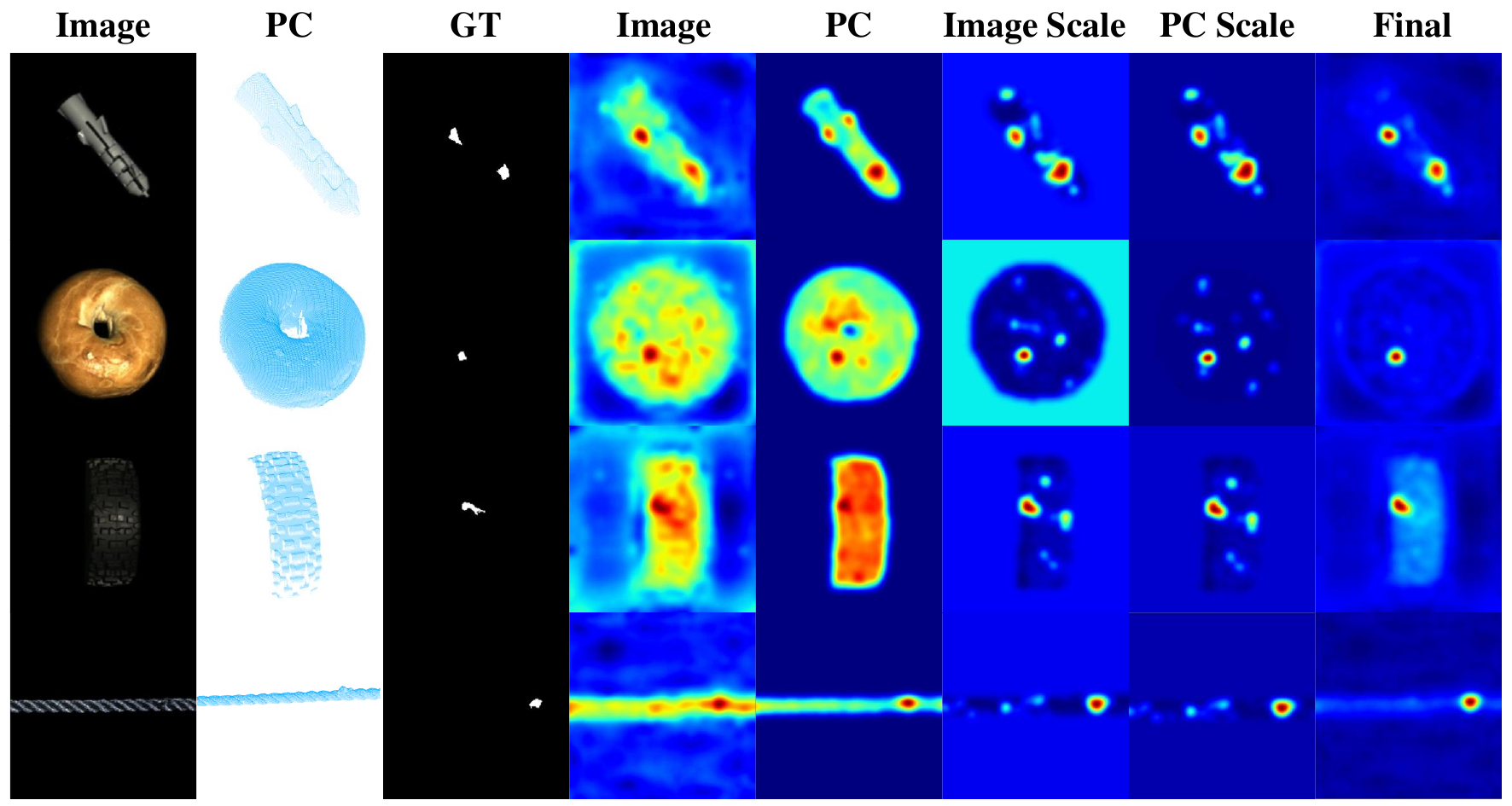}
\caption{Qualitative result on MVTec-3D AD: Original RGB images, 3D point clouds (PC), ground truth (GT), and score maps of $s^{R}_{i}$ (Image), $s^{P}_{i}$ (PC), $w^{R}_{i,0}$ (Image Scale), $w^{P}_{i,0}$ (PC Scale), and $s_{i}$ (Final), from left to right. The scaling factors $(w^{R}_{i,0},w^{P}_{i,0})$ exhibit false positives in certain normal regions, whereas the final score $s_{i}$ effectively suppresses these errors through incorporating distance information from $s^{R}_{i}$ and $s^{P}_{i}$.}
\label{fig:scale}
\end{figure}

\begin{table}[!t]
\centering
\footnotesize
\begin{tabular}{>{\centering\arraybackslash}p{0.6cm}
    >{\centering\arraybackslash}c >{\centering\arraybackslash}c
    >{\centering\arraybackslash}c >{\centering\arraybackslash}c
    } \hline\hline
    \textbf{Score}  & I-AUROC  & P-AUROC &AUPRO@30\% & AUPRO@1\%   \\ \hline
    $s^{R}_i$ & 0.849 & 0.988 & 0.947  & 0.363  \\
    $s^{P}_i$ & 0.854 & 0.975  & 0.915  & 0.319  \\
      $w^{R}_{i,0}$ & 0.916  & 0.948  &  0.916  & 0.439  \\
      $w^{P}_{i,0}$ & 0.915  & 0.984   & 0.958  & 0.443  \\
      $s_{i}$ & \textbf{0.971}  & \textbf{0.997}  & \textbf{0.979}  & \textbf{0.468} \\ \hline\hline
\end{tabular} 
\caption{Analysis of various score metrics on MVTec-3D AD.}
\label{tab:indicator}
\end{table}

\subsection{Comparison on Multimodal Benchmarks}
\label{subsec:comparison}

We illustrate the results of I-AUROC and AUPRO@30\% on MVTec 3D-AD in Table \ref{tab:performance_comparison}, while those of 
AUPRO@1\% and P-AUROC are presented in Table \ref{tab:all_mean}.
Specifically, for anomaly detection, 
the proposed G$^{2}$SF achieves the highest I-AUROC value of 97.1\%. Although LSFA \cite{tu2024self} matches G$^{2}$SF in terms of I-AUROC, its AUPRO@30\% result for anomaly segmentation is 96.8\%, which is lower than G$^{2}$SF's 97.9\%.
Regarding anomaly segmentation, our G$^{2}$SF outperforms all competitors in three pixel metrics 
with a +1.2\% AUPRO@1\% advantage over Shape-guided \cite{Chu2023Shape}. Notably, for AUPRO@30\% value, although Shape-guided is close to G$^{2}$SF, its I-AUROC of 94.7\% is significantly lower than G$^{2}$SF's 97.1\%. Overall, G$^{2}$SF achieves the best performance in both anomaly detection and anomaly segmentation tasks. This performance gain stems from our purpose of learning a discriminative local distance metric. As demonstrated by the qualitative results visualized in Fig. \ref{fig:qualitative}, G$^{2}$SF effectively suppresses overestimated scores in normal regions compared to M3DM \cite{wang2023multimodal} and CFM \cite{costanzino2024multimodal}, while producing sharper anomaly boundaries. To provide a comprehensive visualization,  Fig. \ref{fig:distribution} illustrates pixel-wise anomaly score distributions in four categories, demonstrating that anomaly scores of normal regions in G$^{2}$SF shrinkage significantly within a narrow range. This property effectively reduces false positives, leading to more substantial gains in AUPRO@1\% (Table \ref{tab:all_mean}). In addition, the results on Eyecandies in Table \ref{tab:all_mean} showcase the best performance of our G$^{2}$SF over all metrics, particularly with +2.2\% AUPRO@1\% over CFM \cite{horwitz2023back}. 
More details are provided in the supplementary material.

\subsection{Analysis}
\label{subsec:ablation}

\begin{table}[!t]
\centering
\footnotesize
\begin{tabular}{>{\centering\arraybackslash}p{2.2cm} | >{\centering\arraybackslash}p{0.6cm} |
>{\centering\arraybackslash}p{0.6cm} |
>{\centering\arraybackslash}p{0.6cm} | >{\centering\arraybackslash}p{0.6cm} |
    >{\centering\arraybackslash}p{0.6cm}}
\hline\hline
{\textbf{Components}}&  \multicolumn{5}{c}{\textbf{Stepwise Combinations}}   \\\hline
{$\mathcal{L}_{sep}$} in Eq. \eqref{eq:sep_loss} & \CheckmarkBold  & \CheckmarkBold  & \CheckmarkBold  & \CheckmarkBold  & \CheckmarkBold   \\
 {$\mathcal{L}_{cns}$} in Eq. \eqref{eq:cns_loss} &  \XSolidBrush &\CheckmarkBold   & \CheckmarkBold   & \CheckmarkBold  & \CheckmarkBold   \\
{$\mathcal{L}_{mar}$} in Eq. \eqref{eq:margin_loss} & \XSolidBrush  &  \XSolidBrush & \CheckmarkBold  & \CheckmarkBold   & \CheckmarkBold  \\
{$\mathcal{L}_{sc}$} in Eq. \eqref{eq:sc_loss}& \XSolidBrush  &\XSolidBrush   & \XSolidBrush & \CheckmarkBold  & \CheckmarkBold  \\
{$\mathcal{L}_{cma}$} in Eq. \eqref{eq:cma_loss}&  \XSolidBrush & \XSolidBrush  & \XSolidBrush  & \XSolidBrush  & \CheckmarkBold  \\ \hline
 {I-AUROC} & 0.868 & 0.950  & 0.950 & 0.961  & \textbf{0.971}   \\
{P-AUROC} & 0.994 & 0.995 &  0.996 & \textbf{0.997} &  \textbf{0.997} \\
{AUPRO@30\%} & 0.967 & 0.972 & 0.975 & 0.978  & \textbf{0.979}  \\
\hline\hline
\end {tabular}
\caption{Ablation results of individual loss components within G$^{2}$SF on MVTec-3D AD dataset.}
\label{tab:ablation_loss}
\end {table}

\noindent\textbf{Effectiveness of Final Score} ${s}_i$: Table \ref{tab:indicator} compares key score metrics for anomaly detection: unimodal scores $s^{m}_i$ ($m \in \{P,R\}$) (Euclidean distances, see Section \ref{subsubsec:memory_bank}), scaling factors $w^{m}_{i,0}$ from LSPN in Eq. \eqref{eq:weight}, and our fused score ${s}_i$ in Eq. \eqref{eq:aggregate}. 
As evidenced in Table \ref{tab:indicator}, unimodal anomaly scores $s^{m}_i$ demonstrate suboptimal performance. This limitation stems from the inherent information incompleteness of individual modalities. Our fused score ${s}_i$
overcomes these constraints through crossmodal complementarity, achieving significant improvements over unimodal baselines. These quantitative gains are also supported by Figs. \ref{fig:qualitative} and \ref{fig:scale}.

While $w^{m}_{i,0}$ characterizes the local directions of normal features within local spaces (Section \ref{subsec:lspn}), its exclusion of the valuable distances, i.e., $s^{m}_{i}$, leads to critical limitations. Specifically, proximity to memory prototypes ($s^{m}_i \rightarrow 0$) inherently indicates normality regardless of directional deviations, which is a prior ignored by $w^{m}_{i,0}$. As evidenced in Fig. \ref{fig:scale}, the score maps obtained by $w^{m}_{i,0}$ exhibit noticeable false positives. Using ${s}_i$ in our G$^{2}$SF mitigates this issue, which benefits from sufficiently leveraging complementary directional and distance information.
The comprehensive results are provided in the supplementary material.

\begin{table}[!t]
\centering
\footnotesize
\begin{tabular}{>{\centering\arraybackslash}p{0.6cm}
    >{\centering\arraybackslash}c >{\centering\arraybackslash}c
    >{\centering\arraybackslash}c >{\centering\arraybackslash}c
    } \hline\hline
    \textbf{Strategy}  & I-AUROC  & P-AUROC &AUPRO@30\% & AUPRO@1\%   \\ \hline
    $l_{i,0}$ & 0.963   & \textbf{0.997}   & 0.977    & 0.461  \\
    $\max$  & 0.953   &  0.986   & 0.960   & 0.453    \\
      $\mathrm{mean}$ &  \textbf{0.972}  & 0.983   & 0.955   &   0.458  \\
     $\min$ & 0.971 & \textbf{0.997}  & \textbf{0.979}  & \textbf{0.468} \\ \hline\hline
\end{tabular} 
\caption{Investigation on different score aggregation strategies on MVTec 3D-AD dataset.}
\label{tab:scoring}
\end{table}

\noindent\textbf{Ablation of Loss Components}: Table \ref{tab:ablation_loss} systematically evaluates the contribution of each loss component through stepwise removal. Using the separation loss $\mathcal{L}_{sep}$ as the baseline for the metric learning, the addition of the consistency loss $\mathcal{L}_{cns}$ drives an 8.2\% I-AUROC improvement,
surpassing M3DM \cite{wang2023multimodal} across all metrics and achieving parity with CFM \cite{costanzino2024multimodal}. The margin loss $\mathcal{L}_{mar}$ specifically improves pixel-wise localization (AUPRO@30\% +0.3\%), while the scaling loss $\mathcal{L}_{sc}$ and cross-modal alignment loss $\mathcal{L}_{cma}$ further enhances image-level detection (I-AUROC +1.1/1.0\%). These synergistic effects collectively establish G$^{2}$SF's state-of-the-art performance, validating the necessity of all proposed loss components.

\noindent\textbf{Investigation of Score Aggregation Strategy}: 
Section \ref{subsec:anomaly_scoring} employs a $\min$ operator in Eq. \eqref{eq:aggregate} to aggregate metrics $\{l_{i,j}\}$ from adjacent local spaces. We systematically evaluate three alternatives in Table \ref{tab:scoring}: (1) single-space baseline $l_{i,0}$. (2) $\max$ operator. (3) $\mathrm{mean}$ operator.  The $\min$-based strategy achieves superior overall performance (+0.8\%/0.7\% I-AUROC/AUPRO@1\% vs 
$l_{i,0}$), validating the necessity of multi-space consensus. While $\max$ and $\mathrm{mean}$ operators incorporate adjacent local spaces,
they indiscriminately weight less discriminative $l_{i,j}$ from distant regions, compromising detection precision. The 
$\min$ operator in Eq. \eqref{eq:aggregate} prioritizes the most representative local metric through a proper geometric prior as discussed in Section \ref{subsec:anomaly_scoring}. The results of all categries of MVTec-3D AD are summarized in the supplementary material.

\begin{table}[!t]
\centering
\footnotesize
\begin{tabular}{>{\centering\arraybackslash}p{1.5cm}
    >{\centering\arraybackslash}c >{\centering\arraybackslash}c
    >{\centering\arraybackslash}c >{\centering\arraybackslash}c
    } \hline\hline
    \textbf{Method}  & AST \cite{rudolph2023asymmetric}  & M3DM \cite{wang2023multimodal} & CFM\cite{costanzino2024multimodal} & Ours   \\ \hline
    {\fontsize{4}{1.1}{Frame Rate}}  & 4.97  & 2.12  & \textbf{21.8} & \underline{6.04} \\ \hline\hline
\end{tabular} 
\caption{Frame rate comparison on MVTec 3D-AD dataset.}
\label{tab:fps}
\end{table}

\noindent\textbf{Computational Time}: Table \ref{tab:fps} shows that G$^{2}$SF detects 6.04 samples per second, sufficient for general industrial requirements. 

\noindent\textbf{Analyses of Memory Prototypes and Global Scaling Factors $\sigma_m$}: Supplementary material provides the sensitivity study about the number of memory prototypes, demonstrating relatively stable performance. Additionally, the investigation on the role of $\sigma_m$ is included, showing the necessity of learning optimal $\sigma_m$ instead of fixed ones.

\section{Conclusion}
\label{sec:conclusion}

In this paper, we propose G$^{2}$SF for multimodal anomaly detection, which learns a unified anisotropic local distance metric to replace the indiscriminative isotropic Euclidean metrics. This metric is formulated by the direction-aware scaling factors predicted by LSPN and original Euclidean distances in individual modalities. We propose new loss functions to enable robust metric learning, serving for discrimination or geometric property maintenance. A geometric prior-induced anomaly scoring strategy is proposed to enhance anomaly detection and segmentation. Extensive experimental results show our state-of-art performance compared to existing competitive methods.

\section{Acknowledgement}
\label{sec:Acknowledgement}
This work is supported by the National Natural Science Foundation of China (No. 72371219 and No. 72001139), and the Guangzhou Municipal Science and Technology Project (No. 2025A04J5288). Corresponding author is Juan Du.

{\small
\bibliographystyle{ieeenat_fullname}
\bibliography{main}
}

\clearpage
\setcounter{page}{1}
\maketitlesupplementary

\renewcommand\thesection{\Alph{section}}
\setcounter{section}{0}
\setcounter{figure}{0}
\setcounter{table}{0}

\section{Anomaly Synthesis}
\label{sec:appen_anomaly_synthesis}

\begin{figure}[ht]
\centering 
\includegraphics[width=\linewidth]{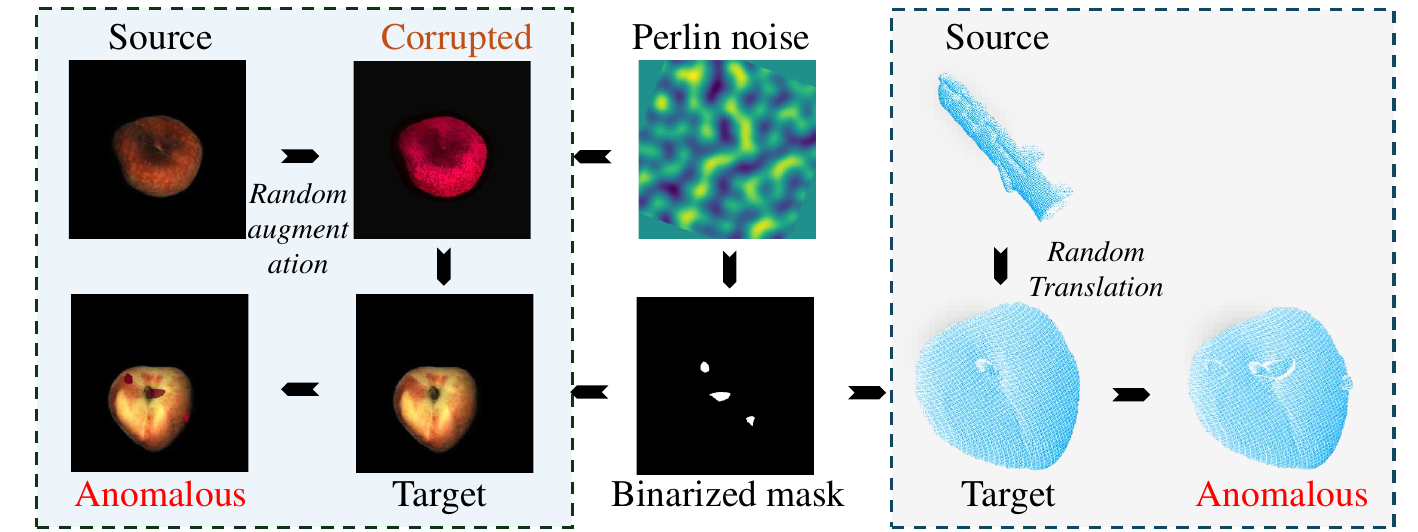}
\caption{Flowchart of the anomaly synthesis procedure based on CutPaste: (1) Berlin noise maps are binarized into masks to select paste locations; (2) Source regions from normal 3D point clouds/RGB images (or generic models) are cut; (3) Random corruptions (pixel/geometry augmentations) are applied; (4) Modified regions are pasted onto target data.}
\label{fig:cut_and_paste}
\end{figure}

The whole procedure of our cutPaste anomaly synthesis is illustrated in Fig. \ref{fig:cut_and_paste}. More generated anomalous RGB images and associated 3D point clouds are provided in Fig. \ref{fig:anomalous_samples}. The synthetic anomalies are pronounced than real-world anomalies in MVTec-3D AD.

\begin{figure}[ht]
\centering 
\includegraphics[width=\linewidth]{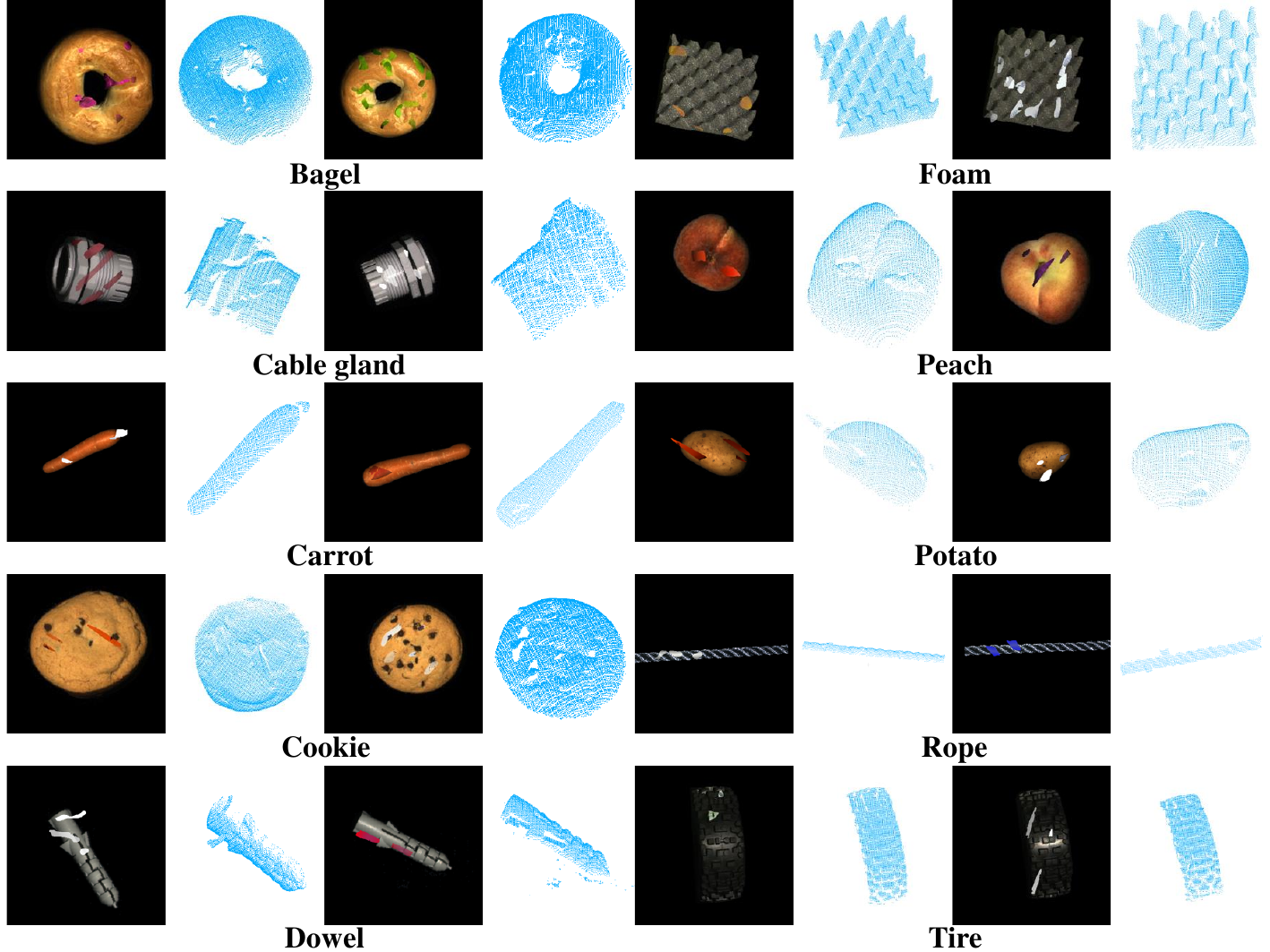}
\caption{Visualization of generated anomalous RGB images and 3D point clouds on MVTec-3D AD.}
\label{fig:anomalous_samples}
\end{figure}

\section{Implementation Details}

\subsection{Feature Extraction}
\label{subsec:feature extraction}
RGB features are extracted using DINO ViT-B/8 \cite{caron2021emerging} pretrained on ImageNet \cite{deng2009imagenet}, generating $28\times28\times768$ feature maps upsampled to $56\times56\times768$. Point clouds are processed via Point-MAE \cite{pang2022masked} pretrained on ShapeNet \cite{chang2015shapenet}, grouping 3D points into 1024 clusters (feature dimension 1152) and interpolated to $56\times56\times1152$ for spatial alignment with image feature maps. We use PatchCore \cite{roth2022towards} to construct modality-specific memory banks with 10\% coreset sampling.  Notably, we consider the original $28\times28\times768$ RGB feature maps instead of the upsampled ones when constructing the memory bank of RGB modality to reduce the memory storage. The above procedure remains the same as done in M3DM \cite{wang2023multimodal}.

\subsection{Data Processing}
To accelerate training and underscore discriminative foreground features,  we pre-collect features from 800 synthetic anomalous samples per class (Section \ref{subsec:anomaly_syn}) (1000 for Eyecandies), forming a feature dataset equally split for training and validation. 
The features originate from foreground regions identified via RANSAC plane fitting on 3D point clouds, where areas exceeding 0.005 distance from the fitted plane are retained following \cite{wang2023multimodal,costanzino2024multimodal}. During evaluation on MVTec 3D-AD, background regions remain included. As background features are excluded during training, we set their scaling factors to unity ($w^{m}_{i,0}=1$) in Eq. \eqref{eq:metric}, serving as a linear combination of the original unimodal results. For Eyecandies, background regions are neglected following CFM \cite{costanzino2024multimodal}.

\begin{table}[!t]
\centering
\footnotesize
\begin{tabular}{>{\centering\arraybackslash}p{2.0cm}
    >{\centering\arraybackslash}c >{\centering\arraybackslash}c
    >{\centering\arraybackslash}c >{\centering\arraybackslash}c >{\centering\arraybackslash}c
    } \hline\hline
    \textbf{Dataset}  & $\alpha$  & $\beta$ & $\gamma$ & $\mu$ & {Batch Size}    \\ \hline
    \textbf{MVTec 3D-AD} &  10   & 60    &   8   &  20  & 8192 \\
    \textbf{Eyecandies}  &  10  &  100   &   2  &  0.5  & 15360  \\
    \hline\hline
\end{tabular} 
\caption{Turning parameters and batch size in our experiment.}
\label{tab:para}
\end{table}

\subsection{Parameter Setting}
For $\mathcal{L}_{cns}$ in Eq. \eqref{eq:cns_loss}, we set $k=5$ and $\eta_{0}=1.2$. We train our LSPN using an Adam optimizer \cite{kingma2014adam} with learning rate $1.5e^{-4}$ and weight decay $1.5e^{-4}$, combined with  $l_1$ regularization of weight $1e^{-4}$ and drop out with ratio 0.5 over 80 epochs.  Global scaling factors $\sigma^m$ are separately optimized at a higher learning rate of $5e^{-3}$ to accelerate convergence. We initialize LSPN by the default strategy in Pytorch, that is, Kaiming Initialization \cite{he2015delving}. We observe $w^{m}_{i,j} \approx 1$ under this strategy, meeting the requirement in Section \ref{subsec:lspn}. For modality $m$, we normalize all Euclidean distances $\{s^m_{i,j}\}$ by their mean over the training dataset. Then, we initialize $\sigma^m=0.5$ with equal importance from each modality. The tuning parameters about loss components and batch size for MVTec 3D-AD and Eyecandies datasets are listed in Table \ref{tab:para}.
Finally, we implement our G$^{2}$SF by Pytorch and run all experiments on Linux with a Nvidia RTX 4090 GPU.

\section{Learning Curve}

\begin{figure}[!t]
\centering 
\includegraphics[width=\linewidth]{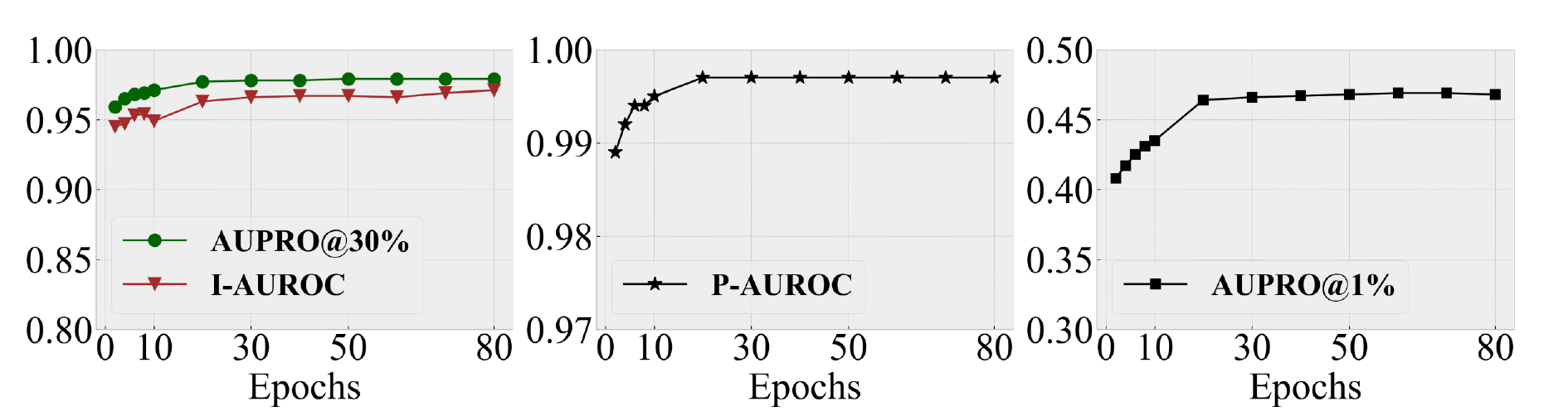}
\caption{Performance over different training epochs on MVTec 3D-AD dataset.}
\label{fig:learning_curve}
\end{figure}

We train multiple models that stop at  \{2,4,6,8,10,20,...,80\} epochs respectively, and report their performances (I-AUROC, P-AUROC, AUPRO@30\%, and AUPRO@1\%) on MVTec-3D AD in Fig. \ref{fig:learning_curve}. 
 After only 4 epochs, G$^{2}$SF outperforms M3DM \cite{wang2023multimodal} (94.5\% I-AUROC, 99.2\%P-AUROC, 96.4\% AUPRO@30\%, and 39.4\%AUPRO@1\%). This benefits from our design, as our model evolves from an established Euclidean metric to the target anisotropic metric, see Section \ref{subsec:lspn}.

\begin{figure}[!t]
\centering 
\includegraphics[width=0.6\linewidth]{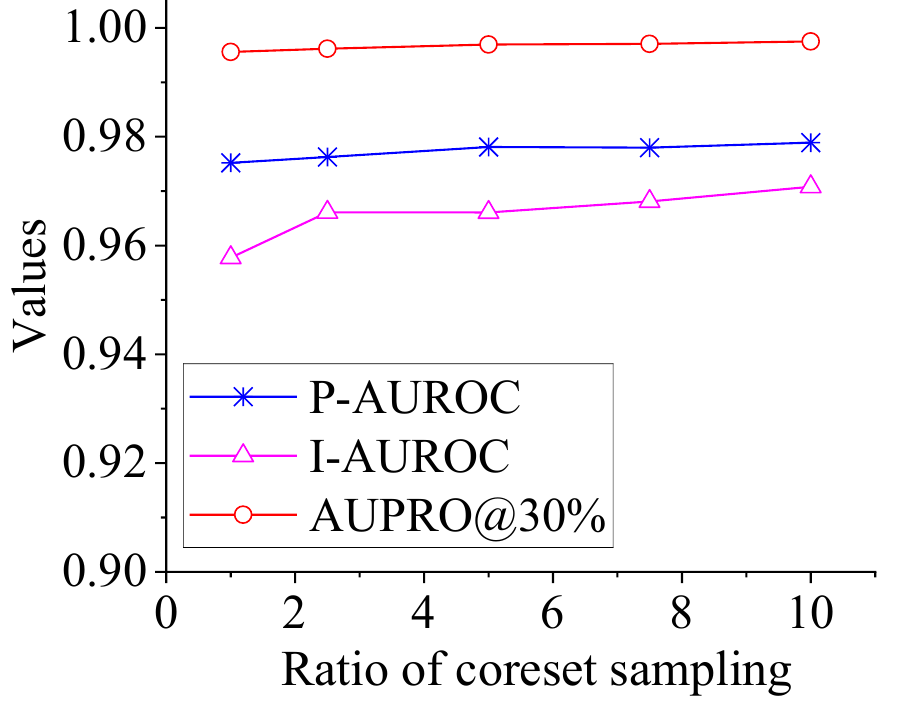}
\caption{Performance over different ratios of coreset sampling on MVTec 3D-AD dataset.}
\label{fig:coreset_ratio}
\end{figure}

\section{Sensitivity Study on Sizes of Memory Banks}
Since the learned metric $l(\cdot)$ is defined in terms of memory prototypes, this subsection presents a sensitivity study of the anomaly detection performance with respect to the number of prototypes. We vary the coreset sampling ratio (see Section \ref{subsec:feature extraction}) at \{10,7.5,5,2.5,1\}\% levels. At the default 10\%, the average numbers of point cloud and RGB image memory prototypes across all categories on the MVTec-3D AD dataset are 83,292 and 20,823, respectively. The corresponding results for the evaluation metrics are summarized in Fig. \ref{fig:coreset_ratio}. At the 2.5 \% level, G$^2$SF achieves 96.6\% I-AUROC, 99.6\% P-AUROC, and 97.6\% AUPRO@30\%. Nevertheless, Fig. \ref{fig:coreset_ratio} demonstrates that the performance of G$^2$SF is relatively stable to the sizes of memory banks.

\begin{table*}[t]
\centering
\footnotesize
\begin{tabular}{
   >{\centering\arraybackslash}m{1.5cm} | c c c c |
    c c c c } \hline\hline

\multirow{2}{*}{\textbf{Metric}}& \multicolumn{4}{c}{\textbf{MVTec-3D AD}} &\multicolumn{4}{c}{\textbf{Eyecandies}} \\ \cline{2-9} 
& I-AUROC & P-AUROC & AUPRO@30\% & AUPRO@1\% & I-AUROC & P-AUROC & AUPRO@30\% & AUPRO@1\% \\ \cline{1-9}
Learn &\textbf{0.971} & \textbf{0.997} & \textbf{0.979} & \textbf{0.468} &\textbf{0.902} & \textbf{0.982} & \textbf{0.898} & \textbf{0.357} \\ 
$\sigma^{P}$=$\sigma^{R}$=0.5 &0.964& 0.994 & 0.972 & 0.460 &0.827 & 0.977 & 0.892 & 0.328  \\\hline\hline

\end{tabular}
\caption{Ablation study of global scaling factors on MVTec-3D AD and Eyecandies datasets.}
\label{tab:global_scale}
\end{table*}

\section{Analysis on Global Scaling Factors}
\begin{figure}[!t]
\centering 
\includegraphics[width=1.0\linewidth]{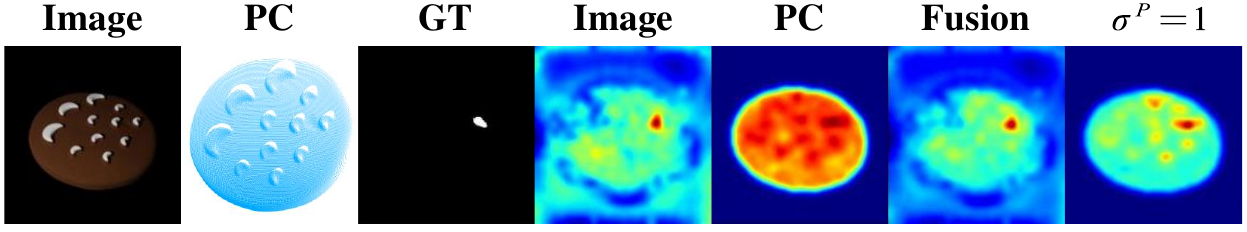}
\caption{Visualization of learned $\sigma^m$ and fixed $\sigma^P=1$ for an example from \textit{ChocolateCookie} category in Eyecandies dataset.}
\label{fig:global_scale}
\end{figure}

G$^2$SF defines trainable global scaling factors $\sigma^m$ ($m\in {P,R}$) in Eq. \eqref{eq:metric} to formulate the metric $l(\cdot)$, where $\sigma^m$ controls the importance of modality $m$ to final $l(\cdot)$  and normalizes the original Euclidean distances. Fig. \ref{fig:global_scale} demonstrates this mechanism prioritizing image modality over less discriminative point cloud modality of the \textit{ChocolateCookie} category in Eyecandies dataset, compared with fusion and image score maps. Moreover, ignoring the contribution of the RGB image modality by setting $\sigma^P=1$ and $\sigma^R=0$ yields significantly degraded results. Furthermore, Table \ref{tab:global_scale} compares performance using learned $\sigma^m$ against fixed $\sigma^P=\sigma^R=0.5$ during inference on both the MVTec 3D-AD and Eyecandies datasets. These results collectively underscore the importance of learning optimal $\sigma^m$ during training.

\section{Results on Eyecandies}

\begin{figure*}[ht]
\centering 
\includegraphics[width=\linewidth]{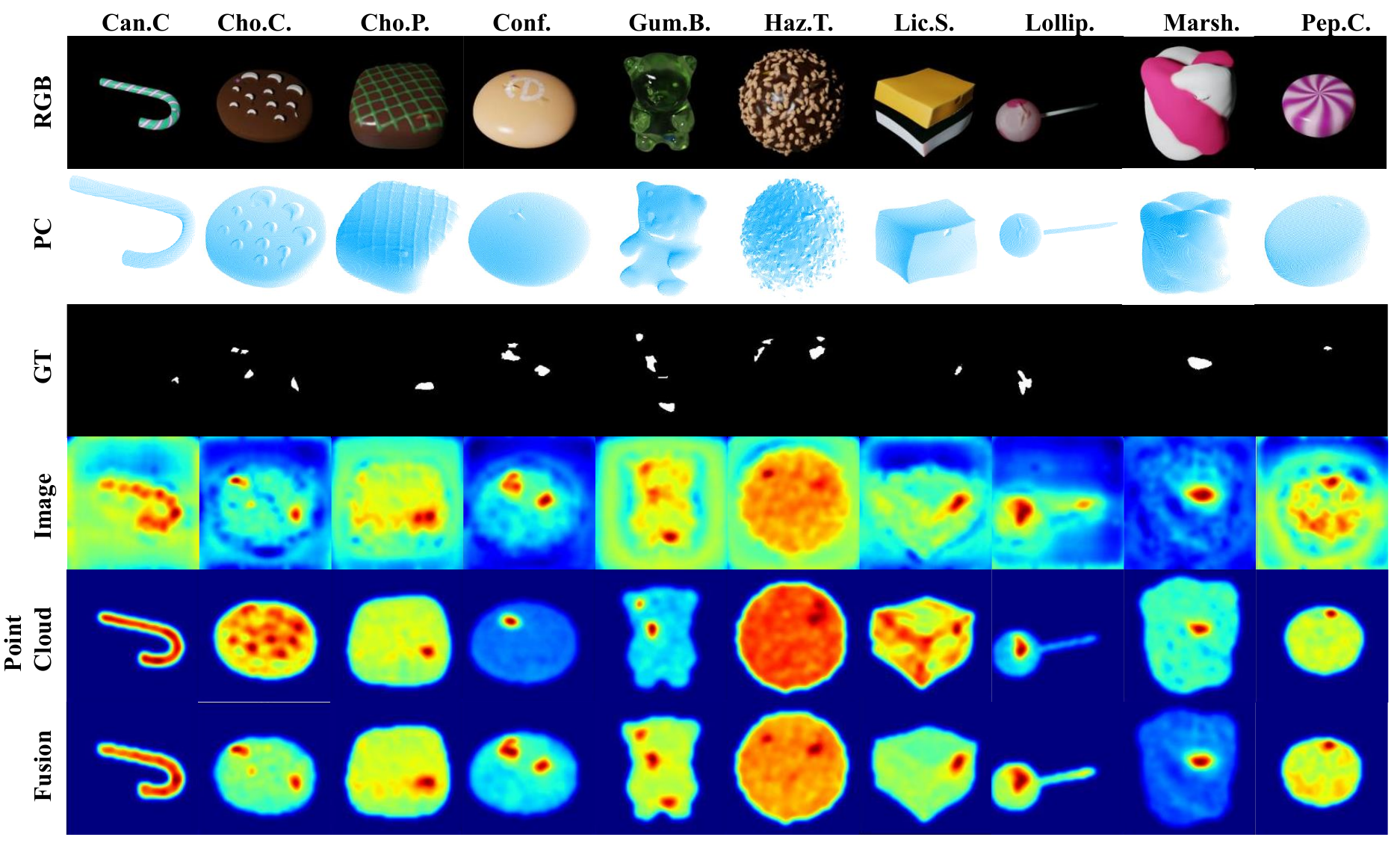}
\caption{Qualitative results on Eyecandies dataset.}
\label{fig:quali_eye}
\end{figure*}

We provide the quantitative results for all categories in Eyecandies in Table \ref{tab:performance_comparison}. G$^2$SF achieves the best performance of all metrics in terms of +0.5\% I-AUROC and +0.8\% AUPRO@30\% over 2M3DF \cite{Asad2M3DF}, +0.5\% P-AUROC over M3DM \cite{wang2023multimodal}, +2.2\%AUPRO@1\%, +2.4\%AUPRO@10\%, and +2.8\%AUPRO@5\% over CFM \cite{costanzino2024multimodal}. The qualitative results are illustrated in Fig. \ref{fig:quali_eye}.

\begin{table*}[ht]
\centering
\footnotesize
\begin{tabular}{>{\centering\arraybackslash}m{0.2cm} |
    >{\centering\arraybackslash}m{0.6cm} >{\centering\arraybackslash}m{0.7cm} |
    >{\centering\arraybackslash}m{0.7cm} >{\centering\arraybackslash}m{0.7cm}
    >{\centering\arraybackslash}m{0.7cm} >{\centering\arraybackslash}m{0.7cm}
    >{\centering\arraybackslash}m{0.7cm} >{\centering\arraybackslash}m{0.7cm}
    >{\centering\arraybackslash}m{0.7cm} >{\centering\arraybackslash}m{0.7cm}
    >{\centering\arraybackslash}m{0.7cm} >{\centering\arraybackslash}m{0.7cm} |
    >{\centering\arraybackslash}m{0.7cm}} \hline\hline
     & \multicolumn{2}{c|}{\textbf{Method}} & \textit{Can. C.}& \textit{Cho. C.} & \textit{Cho. P.} & \textit{Conf.} & \textit{Gum. B.} & \textit{Haz. T.} & \textit{Lic. S.} & \textit{Lollip.} & \textit{Marsh.} & \textit{Pep. C.} & \textbf{Mean} \\ \hline
     
   \multirow{5}{*}{\rotatebox{90}{\textbf{I-AUROC}}} 
 & \multicolumn{2}{c|}{AST\cite{rudolph2023asymmetric}} & 0.574& 0.747& 0.747& 0.889 &0.596& 0.617& 0.816& 0.841 &0.987& 0.987& 0.780 \\  
 & \multicolumn{2}{c|}{M3DM* \cite{wang2023multimodal}} &0.597& 0.954& 0.931 &0.990 &0.883 &0.666 &0.923& \underline{0.888} &0.995& \textbf{1.000}& 0.882 \\  
 & \multicolumn{2}{c|}{CFM \cite{costanzino2024multimodal}} &0.680& 0.931& 0.952& 0.880 &0.865 &\underline{0.782}& 0.917& 0.840 &\underline{0.998}& 0.962& 0.881\\
 & \multicolumn{2}{c|}{2M3DF \cite{Asad2M3DF}} &\textbf{0.753}& 0.955& 0.937& 0.967& \underline{0.901} & \textbf{0.792}& 0.889& \textbf{0.913}& 0.980& 0.893& 0.897\\ 
& \multicolumn{2}{c|}{Ours} & \underline{0.696} & \textbf{0.963} & \textbf{0.967} & \textbf{1.000} & \textbf{0.907} &0.701 & \underline{0.939} &0.855 &0.989 & \textbf{1.000} & \textbf{0.902} \\ 
\hline\hline

\multirow{6}{*}{\rotatebox{90}{\textbf{P-AUROC}}}
 & \multicolumn{2}{c|}{AST\cite{rudolph2023asymmetric}} &0.763& 0.960& 0.911& 0.969 &0.788 &0.837 &0.918 &0.924 &0.983 &0.968 &0.902 \\  
 & \multicolumn{2}{c|}{M3DM* \cite{wang2023multimodal}} &0.968 &\underline{0.986} &0.964& \textbf{0.998}& 0.976& 0.928 &\underline{0.976} &0.988& 0.996& \underline{0.995}& \underline{0.977} \\ 
 & \multicolumn{2}{c|}{CFM \cite{costanzino2024multimodal}} &\textbf{0.985}& 0.984& 0.961 &0.986 &0.958 &0.937 &0.968 &0.981 &0.994& 0.978& 0.973 \\
 & \multicolumn{2}{c|}{LSFA \cite{tu2024self}} &0.969 &0.957 &\underline{0.967} &\underline{0.996} &0.971 &\underline{0.938} &0.970 &\underline{0.990}& 0.998 &\textbf{0.987}& 0.974 \\
& \multicolumn{2}{c|}{Ours} & \underline{0.981} &0.982&\textbf{0.977}&\textbf{0.998}&\textbf{0.982}&0.936&\textbf{0.981}&\underline{0.989}&\underline{0.997}&\textbf{0.998}&\textbf{0.982} \\ \hline\hline

\multirow{5}{*}{\rotatebox{90}{\textbf{AUPRO@30\%}}}
 & \multicolumn{2}{c|}{AST\cite{rudolph2023asymmetric}} &0.514 &0.835& 0.714 &0.905& 0.587 &0.590 &0.736 &0.769 &0.918 &0.878 &0.744 \\  
 & \multicolumn{2}{c|}{M3DM* \cite{wang2023multimodal}} &0.889 &0.921 &0.808 & \underline{0.982} &\underline{0.889} &0.675 &0.872 &0.901 &0.964 & \underline{0.973} & 0.887 \\ 
 & \multicolumn{2}{c|}{CFM \cite{costanzino2024multimodal}} &\textbf{0.942} &0.902 &0.831 &0.965 &0.875 &\underline{0.762} &0.791& \underline{0.913} & 0.939 &0.949& 0.887 \\
 & \multicolumn{2}{c|}{2M3DF \cite{Asad2M3DF}} &0.924 &\textbf{0.935} &\underline{0.820} &0.940& 0.875& \textbf{0.781} &0.816 &\textbf{0.923} &0.958 &0.926 & \underline{0.890} \\
& \multicolumn{2}{c|}{Ours} & \underline{0.928} & 0.897 &\textbf{0.843} &\underline{0.982} &\textbf{0.890}&0.687& \textbf{0.888} & \underline{0.913} &\textbf{0.971} &\textbf{0.981} & \textbf{0.898} \\ \hline\hline

\multirow{4}{*}{\rotatebox{90}{\tiny{\textbf{AUPRO@1\%}}}}
 & \multicolumn{2}{c|}{AST\cite{rudolph2023asymmetric}} &0.035 &0.230 &0.129 &0.234 &0.092 &0.069 &0.139 &0.090 &0.255& 0.224& 0.149 \\  
 & \multicolumn{2}{c|}{M3DM* \cite{wang2023multimodal}} &0.166 &0.388 &0.329 &\underline{0.486} &0.315 &0.131 &0.323 &\underline{0.258} & \underline{0.462} & \underline{0.454} & 0.331 \\ 
 & \multicolumn{2}{c|}{CFM \cite{costanzino2024multimodal}} &\textbf{0.229} &\underline{0.397} &\underline{0.345} &0.389 &\underline{0.353} & \textbf{0.188} & \underline{0.333} &0.236 &0.455 &0.428& \underline{0.335} \\
& \multicolumn{2}{c|}{Ours} &\underline{0.174}&\textbf{0.416}&\textbf{0.377}&\textbf{0.487}&\textbf{0.360}&\underline{0.164}&\textbf{0.353}&\textbf{0.281}&\textbf{0.478}&\textbf{0.479}& \textbf{0.357} \\ \hline\hline

\multirow{4}{*}{\rotatebox{90}{\tiny{\textbf{AUPRO@10\%}}}}
 & \multicolumn{2}{c|}{AST\cite{rudolph2023asymmetric}}  &0.285 &0.709& 0.545 &0.770& 0.404& 0.350 &0.584& 0.544& 0.770 &0.744 &0.570 \\  
 & \multicolumn{2}{c|}{M3DM* \cite{wang2023multimodal}} & 0.677& 0.836& 0.698& \textbf{0.947} & \underline{0.754} & 0.410 & \underline{0.732} & \underline{0.712} & \underline{0.913} & \underline{0.924} &0.760 \\ 
 & \multicolumn{2}{c|}{CFM \cite{costanzino2024multimodal}} & \textbf{0.827} &0.815 &0.731 &0.896 &0.741 &\textbf{0.550} &0.663& \textbf{0.739} &0.893& 0.868 & \underline{0.772} \\
& \multicolumn{2}{c|}{Ours} & \underline{0.784} & \textbf{0.820} & \textbf{0.752} & \underline{0.946} & \textbf{0.780} & \underline{0.490} & \textbf{0.787} & \textbf{0.739} &\textbf{0.922} & \textbf{0.943} & \textbf{0.796} \\ \hline\hline

\multirow{4}{*}{\rotatebox{90}{\tiny{\textbf{AUPRO@5\%}}}}
 & \multicolumn{2}{c|}{AST\cite{rudolph2023asymmetric}}  &0.173 & 0.592 &0.421& 0.635 &0.288 &0.242 &0.461& 0.378& 0.634& 0.617 &0.444 \\  
 & \multicolumn{2}{c|}{M3DM* \cite{wang2023multimodal}} &0.479 &\textbf{0.759} &0.626 & \textbf{0.894} &0.655 &0.300 & \underline{0.634} & \underline{0.562} & \underline{0.849} &\underline{0.861} & 0.661 \\ 
 & \multicolumn{2}{c|}{CFM \cite{costanzino2024multimodal}} & \textbf{0.662} & \underline{0.750} &\underline{0.653} & 0.801 & \underline{0.657} & \textbf{0.427} &0.609& 0.552 &0.838& 0.796& \underline{0.675} \\
& \multicolumn{2}{c|}{Ours} & \underline{0.578} & \textbf{0.759} & \textbf{0.687} & \underline{0.892} & \textbf{0.703} & \underline{0.399} & \textbf{0.698} & \textbf{0.568} & \textbf{0.858} & \textbf{0.889} & \textbf{0.703} \\ \hline\hline

\end{tabular}
\caption{Results on Eyecandies. Best results in \textbf{bold}, runner-ups \underline{underlined}.}
\label{tab:eyecandies}
\end{table*}

\section{Additional Results on MVTec-3D AD}

We summarize the results of different score metrics for all categories in MVTec-3D AD in Table \ref{tab:additional_score}. The results demonstrate that the final score $s_i$ outperforms unimodal scores $\{s^{m}_{i,0}\}$ and scaling factors $\{w^{m}_{i,0}\}$ almost all categories.
Additionally, comprehensive quantitative results of various score aggregation operators are provided in \ref{tab:additional_aggregation}, which validate the superior performance of $\min$ operator over the majority of classes, owing to its geometric interpretation, as discussed in Section \ref{subsec:anomaly_scoring}.

Furthermore, Fig. \ref{fig:anomalous_samples} provides additional qualitative results for all classes in MVTec-3D AD. Consistent with Fig. \ref{fig:qualitative}, our G$^{2}$SF suppresses anomaly scores of normal regions and provides clearer anomaly boundaries, compared to M3DM \cite{wang2023multimodal} and CFM \cite{costanzino2024multimodal}. 

\begin{table*}[ht]
\centering
\footnotesize

\begin{tabular}{>{\centering\arraybackslash}m{0.2cm} |
    >{\centering\arraybackslash}m{0.6cm} >{\centering\arraybackslash}m{0.6cm} |
    >{\centering\arraybackslash}m{0.6cm} >{\centering\arraybackslash}m{0.6cm}
    >{\centering\arraybackslash}m{0.6cm} >{\centering\arraybackslash}m{0.6cm}
    >{\centering\arraybackslash}m{0.6cm} >{\centering\arraybackslash}m{0.6cm}
    >{\centering\arraybackslash}m{0.6cm} >{\centering\arraybackslash}m{0.6cm}
    >{\centering\arraybackslash}m{0.6cm} >{\centering\arraybackslash}m{0.6cm} |
    >{\centering\arraybackslash}m{0.6cm}} \hline\hline
     & \multicolumn{2}{c|}{\textbf{Method}}  & \textit{Bagel} & \textit{Cable Gland} & \textit{Carrot} & \textit{Cookie} & \textit{Dowel} & \textit{Foam} & \textit{Peach} & \textit{Potato} & \textit{Rope} & \textit{Tire} & \textbf{Mean} \\ \hline

   \multirow{5}{*}{\rotatebox{90}{\textbf{AUPRO@1\%}}} 
   & \multicolumn{2}{c|}{BTF\cite{horwitz2023back}} &0.428 &0.365 &0.452& 0.431 &0.370& 0.244 &0.427& 0.470& 0.298& 0.345& 0.383 \\  
 & \multicolumn{2}{c|}{AST\cite{rudolph2023asymmetric}} &0.388 &0.322& 0.470& 0.411& 0.328 &0.275& \underline{0.474} & \underline{0.487}& 0.360& \underline{0.474} & 0.398 \\  
  & \multicolumn{2}{c|}{Shape-guided\cite{Chu2023Shape}} &- &-& -& -& - &-& - & -& -& - & \underline{0.456}\\ 
 & \multicolumn{2}{c|}{M3DM \cite{wang2023multimodal}} &0.414 &0.395 &0.447& 0.318& \textbf{0.422} &0.335 &0.444& 0.351 &0.416& 0.398& 0.394 \\  
 & \multicolumn{2}{c|}{CFM \cite{costanzino2024multimodal}} &\underline{0.459}& \underline{0.431} & \textbf{0.485} & \underline{0.469} &0.394& \underline{0.413} & 0.468 & \underline{0.487} & \underline{0.464} & \textbf{0.476} & 0.455\\
& \multicolumn{2}{c|}{Ours} &\textbf{0.481} & \textbf{0.443} & \underline{0.484}& \textbf{0.471} & \underline{0.410} & \textbf{0.468} & \textbf{0.487} & \textbf{0.499} & \textbf{0.468}& 0.471& \textbf{0.468}\\ 
\hline\hline

   \multirow{2}{*}{\rotatebox{90}{\tiny\textbf{.@10\%}}} 
 & \multicolumn{2}{c|}{CFM \cite{costanzino2024multimodal}} &0.937 &0.917& 0.947& 0.897& 0.855& 0.906 &0.942 &0.947 &0.926& \textbf{0.944} & \underline{0.922} \\
& \multicolumn{2}{c|}{Ours} &\textbf{0.946} & \textbf{0.928} & \textbf{0.947} & \textbf{0.937} & \textbf{0.912} & \textbf{0.932} & \textbf{0.946} & \textbf{0.950} & \textbf{0.933} & \underline{0.942} & \textbf{0.937}\\ 
\hline\hline

   \multirow{2}{*}{\rotatebox{90}{\tiny\textbf{.@5\%}}} 
 & \multicolumn{2}{c|}{CFM \cite{costanzino2024multimodal}} &\underline{0.877}& \underline{0.843}& \textbf{0.894} & \underline{0.840} & \underline{0.765} &\underline{0.828} & \underline{0.884} & \underline{0.894} & \underline{0.865} & \textbf{0.889} & \underline{0.858} \\
& \multicolumn{2}{c|}{Ours} &\textbf{0.892} &\textbf{0.860} &\underline{0.893} & \textbf{0.877} & \textbf{0.828} &\textbf{0.873} &\textbf{0.892} &\textbf{0.899} &\textbf{0.870} & \underline{0.885} & \textbf{0.877}\\ 
\hline\hline

\end{tabular}
\caption{Additional \textbf{AUPRO@1\%}, \textbf{AUPRO@10\%} and \textbf{AUPRO@5\%} results on MVTec-3D AD. Best results in \textbf{bold}, runner-ups \underline{underlined}.}
\label{tab:performance_comparison}
\end{table*}

\newpage

\begin{table*}[ht]
\centering
\footnotesize

\begin{tabular}{>{\centering\arraybackslash}m{0.2cm} |
    >{\centering\arraybackslash}m{0.6cm} >{\centering\arraybackslash}m{0.6cm} |
    >{\centering\arraybackslash}m{0.6cm} >{\centering\arraybackslash}m{0.6cm}
    >{\centering\arraybackslash}m{0.6cm} >{\centering\arraybackslash}m{0.6cm}
    >{\centering\arraybackslash}m{0.6cm} >{\centering\arraybackslash}m{0.6cm}
    >{\centering\arraybackslash}m{0.6cm} >{\centering\arraybackslash}m{0.6cm}
    >{\centering\arraybackslash}m{0.6cm} >{\centering\arraybackslash}m{0.6cm} |
    >{\centering\arraybackslash}m{0.6cm}} \hline\hline
     & \multicolumn{2}{c|}{\textbf{Method}}  & \textit{Bagel} & \textit{Cable Gland} & \textit{Carrot} & \textit{Cookie} & \textit{Dowel} & \textit{Foam} & \textit{Peach} & \textit{Potato} & \textit{Rope} & \textit{Tire} & \textbf{Mean} \\ \hline
     
   \multirow{5}{*}{\rotatebox{90}{\footnotesize\textbf{I-AUROC}}} 
 & \multicolumn{2}{c|}{ $s^{R}_{i,0}$} &0.945 & \textbf{0.939} & 0.914 & 0.733 & \underline{0.939} & 0.768 &0.944 & 0.621 & 0.930 & 0.755 & 0.849 \\
& \multicolumn{2}{c|}{ $s^{P}_{i,0}$} &0.951 & 0.637 &0.981 &\underline{0.952} &0.837 &0.786 &0.923 &0.930 &0.858 &0.688 &0.854 \\
& \multicolumn{2}{c|}{ $w^{R}_{i,0}$} &\underline{0.969} &0.732 &\underline{0.989} &0.928 &0.831 &\underline{0.983} &0.981 &\textbf{0.988} &\underline{0.965} &\underline{0.798} &\underline{0.916} \\ 
& \multicolumn{2}{c|}{$w^{P}_{i,0}$}&\underline{0.969} &0.727 &\underline{0.989} &0.930 &0.829 &\underline{0.983} &\underline{0.982} &\textbf{0.988} &0.963 &0.796 &0.915 \\
& \multicolumn{2}{c|}{$s_{i}$} & \textbf{0.997} & \underline{0.923} & \textbf{0.993} & \textbf{0.967}& \textbf{0.966}& \textbf{0.991}& \textbf{0.994}& \textbf{0.988}& \textbf{0.966} &\textbf{0.922} & \textbf{0.971} \\ 

\hline\hline

   \multirow{5}{*}{\rotatebox{90}{\footnotesize\textbf{P-AUROC}}} 
 & \multicolumn{2}{c|}{ $s^{R}_{i,0}$} & 0.992  & \underline{0.993}  & 0.995  & \underline{0.976}  & \underline{0.996}  & 0.955  & 0.994  & 0.991  & 0.995  & \underline{0.995}  & \underline{0.988}  \\
& \multicolumn{2}{c|}{ $s^{P}_{i,0}$} &0.987  & 0.945  & 0.997  & 0.940  & 0.981  & 0.935  & 0.993  & 0.996  & 0.994  & 0.981  & 0.975 \\
& \multicolumn{2}{c|}{ $w^{R}_{i,0}$}& 0.964  & 0.910  & 0.995  & 0.774  & 0.954  & 0.973  & 0.965  & 0.998  & 0.997  & 0.952  & 0.948 \\ 
& \multicolumn{2}{c|}{$w^{P}_{i,0}$} &\underline{0.993}  & 0.983  & \underline{0.998}  & 0.916  & 0.986  & \underline{0.991}  & \underline{0.995}  & \textbf{0.999 } & \underline{0.998}  & 0.977  & 0.984 \\
& \multicolumn{2}{c|}{$s_{i}$} & \textbf{0.998 } & \textbf{0.995 } & \textbf{0.999 } & \textbf{0.996 } & \textbf{0.996 } & \textbf{0.997 } & \textbf{0.998} & \textbf{0.999 } & \textbf{0.999 } & \textbf{0.998} & \textbf{0.997}  \\
\hline\hline

   \multirow{5}{*}{\rotatebox{90}{\footnotesize\textbf{AUPRO@30\%}}} 
 & \multicolumn{2}{c|}{ $s^{R}_{i,0}$}& 0.954 & \underline{0.973} & 0.974 & 0.887 & \textbf{0.974} & 0.847 & 0.971 & 0.957 & 0.965 & \underline{0.970}  & 0.947 \\
& \multicolumn{2}{c|}{ $s^{P}_{i,0}$}& 0.96  & 0.806 & 0.977 & 0.899 & 0.929 & 0.761 & 0.971 & 0.976 & 0.946 & 0.927 & 0.915 \\
& \multicolumn{2}{c|}{ $w^{R}_{i,0}$}& 0.957 & 0.900   & 0.976 & 0.724 & 0.896 & 0.911 & 0.944 & 0.982 & 0.974 & 0.898 & 0.916  \\ 
& \multicolumn{2}{c|}{$w^{P}_{i,0}$}& \underline{0.979} & 0.953 & \underline{0.981} & \underline{0.908} & 0.934 & \underline{0.950}  & \underline{0.978} & \textbf{0.983} & \underline{0.976} & 0.939 & \underline{0.958} \\
& \multicolumn{2}{c|}{$s_{i}$} & \textbf{0.982} & \textbf{0.976} & \textbf{0.982} & \textbf{0.979} & \underline{0.971} & \textbf{0.976} & \textbf{0.982} & \textbf{0.983} & \textbf{0.978} & \textbf{0.981} & \textbf{0.979} \\ 

\hline\hline

   \multirow{5}{*}{\rotatebox{90}{\footnotesize\textbf{AUPRO@1\%}}} & \multicolumn{2}{c|}{ $s^{P}_{i,0}$} 
 & 0.367  & \underline{0.423}  & 0.416  & 0.235  & \textbf{0.415 } & 0.296  & 0.399  & 0.285  & 0.396  & 0.399  & 0.363  \\
& \multicolumn{2}{c|}{ $s^{P}_{i,0}$} & 0.424  & 0.063  & 0.448  & 0.406  & 0.243  & 0.238  & 0.385  & 0.409  & 0.356  & 0.222  & 0.319  \\
& \multicolumn{2}{c|}{ $w^{R}_{i,0}$}& 0.480  & 0.384  & 0.480  & 0.378  & 0.356  & \underline{0.469}  & 0.477  & 0.494  & \underline{0.455}  & 0.416  & 0.439  \\ 
& \multicolumn{2}{c|}{$w^{P}_{i,0}$} &\textbf{0.481 } & 0.378  & \underline{0.481}  & \underline{0.421}  & 0.356  & \textbf{0.474 } & \underline{0.480}  & \underline{0.494}  & 0.453  & 0.415  & 0.443 \\
& \multicolumn{2}{c|}{$s_{i}$} & \textbf{0.481 } & \textbf{0.443 } & \textbf{0.484 } & \textbf{0.471 } & \underline{0.410}  & 0.468  & \textbf{0.487 } & \textbf{0.499 } & \textbf{0.468 } & \textbf{0.471 } & \textbf{0.468} \\ 

\hline\hline
\end{tabular}
\caption{Results of various score metrics on MVTec-3D AD. Best results in \textbf{bold}, runner-ups \underline{underlined}.}
\label{tab:additional_score}
\end{table*}

\begin{table*}[ht]
\centering
\footnotesize

\begin{tabular}{>{\centering\arraybackslash}m{0.2cm} |
    >{\centering\arraybackslash}m{0.6cm} >{\centering\arraybackslash}m{0.6cm} |
    >{\centering\arraybackslash}m{0.6cm} >{\centering\arraybackslash}m{0.6cm}
    >{\centering\arraybackslash}m{0.6cm} >{\centering\arraybackslash}m{0.6cm}
    >{\centering\arraybackslash}m{0.6cm} >{\centering\arraybackslash}m{0.6cm}
    >{\centering\arraybackslash}m{0.6cm} >{\centering\arraybackslash}m{0.6cm}
    >{\centering\arraybackslash}m{0.6cm} >{\centering\arraybackslash}m{0.6cm} |
    >{\centering\arraybackslash}m{0.6cm}} \hline\hline
     & \multicolumn{2}{c|}{\textbf{Method}}  & \textit{Bagel} & \textit{Cable Gland} & \textit{Carrot} & \textit{Cookie} & \textit{Dowel} & \textit{Foam} & \textit{Peach} & \textit{Potato} & \textit{Rope} & \textit{Tire} & \textbf{Mean} \\ \hline
     
   \multirow{4}{*}{\rotatebox{90}{\tiny\textbf{I-AUROC}}} 
 & \multicolumn{2}{c|}{ $l_{i,0}$} &\underline{0.997} &	0.925 &	\underline{0.989} &	0.962 &	0.953 &	\textbf{0.993} & 	\underline{0.996} 	&0.964 &	\underline{0.960} &	0.892 &	0.963  \\
& \multicolumn{2}{c|}{ $\max$} &\textbf{0.998} &	\underline{0.932}& 	0.982 	&0.966 	&0.949 	&0.968 &	\textbf{0.999} 	&0.862 &	0.953 &	\underline{0.922} 	&0.953 \\
& \multicolumn{2}{c|}{ $\mathrm{mean}$} &\underline{0.997}& \textbf{0.940}& 	0.988& 	\textbf{0.975} &	\underline{0.954} &	\textbf{0.993} &	0.995& 	\underline{0.980} &	\underline{0.960} 	& \textbf{0.941} &\textbf{0.972} \\ 
& \multicolumn{2}{c|}{$\min$} & \underline{0.997} & 0.923 & \textbf{0.993} & \underline{0.967}& \textbf{0.966}& 0.991& 0.994& \textbf{0.988}& \textbf{0.966} &\underline{0.922} & \underline{0.971} \\ 

\hline\hline

   \multirow{4}{*}{\rotatebox{90}{\tiny\textbf{P-AUROC}}} 
 & \multicolumn{2}{c|}{ $l_{i,0}$} &\underline{0.997} &	\underline{0.993} &	\textbf{0.999} &	\underline{0.994} &	\underline{0.995} &	\textbf{0.997} &	\textbf{0.998} &	\textbf{0.999} &	\textbf{0.999} &	0.997 &	\textbf{0.997}  \\
& \multicolumn{2}{c|}{ $\max$}  &0.980 &	0.989 &	\textbf{0.999} &	0.917 &	0.993 &	0.986 &	\textbf{0.998} 	&\textbf{0.999} &	0.998 	&0.996 &	0.986 \\
& \multicolumn{2}{c|}{ $\mathrm{mean}$}& 0.977 &	0.992 &	\textbf{0.999} 	&0.888 	& \underline{0.995} &	0.983 &	0.997 	&\textbf{0.999} &	\textbf{0.999} &	\textbf{0.998} 	&0.983  \\ 
& \multicolumn{2}{c|}{$\min$} & \textbf{0.998 } & \textbf{0.995 } & \textbf{0.999 } & \textbf{0.996 } & \textbf{0.996 } & \textbf{0.997 } & \textbf{0.998} & \textbf{0.999 } & \textbf{0.999 } & \textbf{0.998} & \textbf{0.997}  \\
\hline\hline

   \multirow{4}{*}{\rotatebox{90}{\tiny\textbf{AUPRO@30\%}}} 
 & \multicolumn{2}{c|}{ $l_{i,0}$} &\underline{0.981} &	\underline{0.970} &	\textbf{0.982} &	\underline{0.974} &	\underline{0.967} &	\underline{0.975} &	\textbf{0.982} 	&\textbf{0.983} &\textbf{0.978} &	\underline{0.979} &	\underline{0.977} \\
& \multicolumn{2}{c|}{ $\max$} & 0.968 	&0.960 &	\textbf{0.982} &	0.883 &	0.954 &	0.936 &	\textbf{0.982} 	&\textbf{0.983} &0.976 &	0.974 &	0.960 \\
& \multicolumn{2}{c|}{ $\mathrm{mean}$} &0.965 &	\underline{0.970} &	\textbf{0.982}&	0.815 &	0.961 &	0.932 	&\textbf{0.982} &	\textbf{0.983} 	&\textbf{0.978} &\underline{0.979}& 	0.955  \\ 
& \multicolumn{2}{c|}{$\min$} & \textbf{0.982} & \textbf{0.976} & \textbf{0.982} & \textbf{0.979} & \textbf{0.971} & \textbf{0.976} & \textbf{0.982} & \textbf{0.983} & \textbf{0.978} & \textbf{0.981} & \textbf{0.979} \\ 

\hline\hline

   \multirow{4}{*}{\rotatebox{90}{\tiny\textbf{AUPRO@1\%}}} & \multicolumn{2}{c|}{$l_{i,0}$} 
 &0.480 &0.429 &	0.482 &	\underline{0.444}& 	\textbf{0.411} &	0.467 &	0.485 &	\underline{0.497} &	\underline{0.465} &0.448 & \underline{0.461} \\
& \multicolumn{2}{c|}{ $\max$} & \textbf{0.481 } &	0.390 &	\textbf{0.485} &	0.428 &	0.391 &	0.467 	&\textbf{0.494} &	0.495 &	0.457 &	0.443 &	0.453 \\
& \multicolumn{2}{c|}{ $\mathrm{mean}$}& 0.480 &	\underline{0.436} &	\textbf{0.485} &	0.396 &	0.403 &	\textbf{0.469} 	&\underline{0.491} &\underline{0.497} &	\underline{0.465} &	\underline{0.460} &	0.458   \\ 
& \multicolumn{2}{c|}{$\min$} & \textbf{0.481 } & \textbf{0.443 } & 0.484 & \textbf{0.471} & \underline{0.410}  & \underline{0.468}  & 0.487 & \textbf{0.499 } & \textbf{0.468 } & \textbf{0.471 } & \textbf{0.468} \\ 

\hline\hline
\end{tabular}
\caption{Results of various score aggregation strategies on MVTec-3D AD. Best results in \textbf{bold}, runner-ups \underline{underlined}.}
\label{tab:additional_aggregation}
\end{table*}

\begin{figure*}[ht]
\centering 
\includegraphics[width=\linewidth]{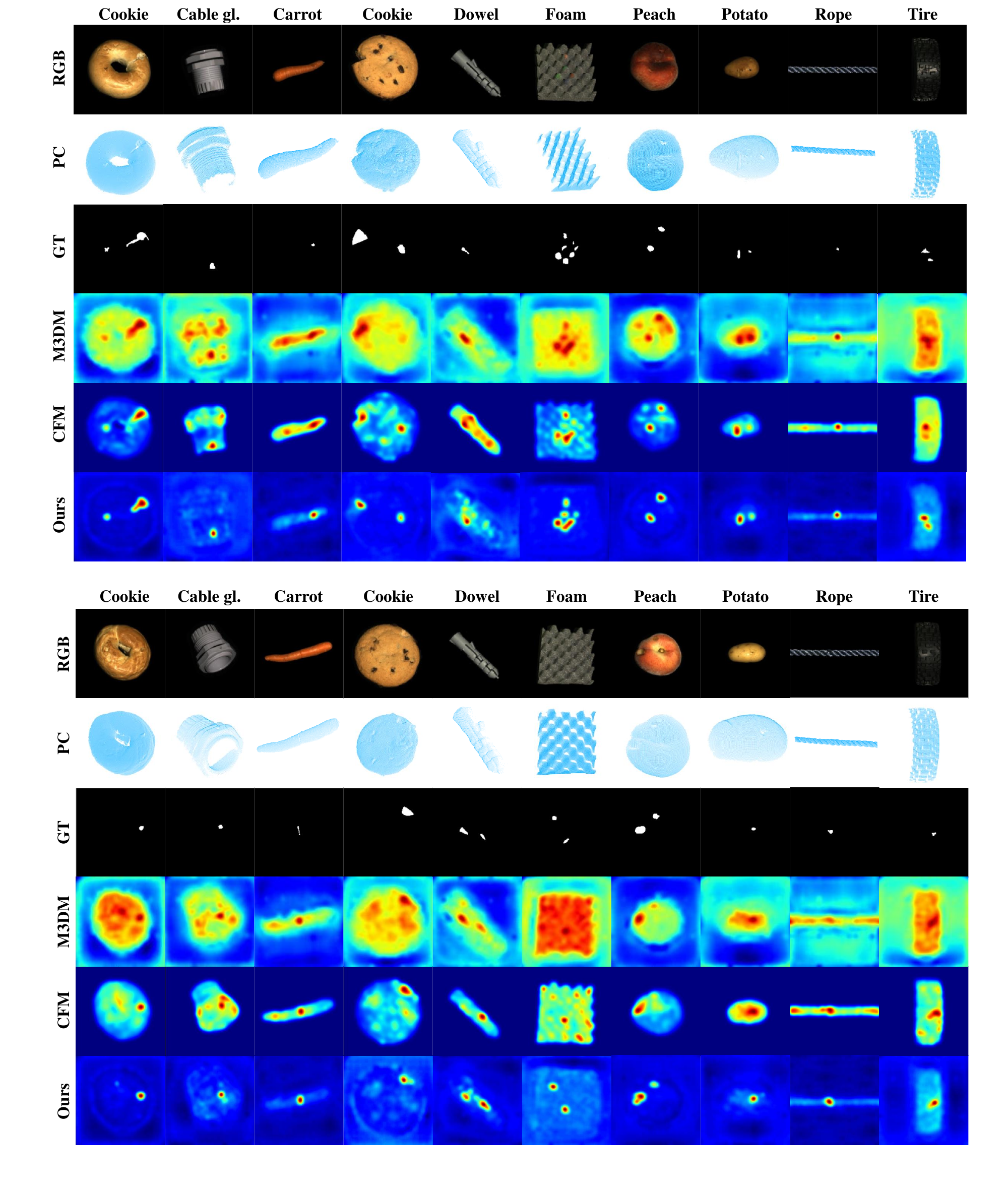}
\caption{Additional qualitative results on MVTec-3D AD dataset.}
\label{fig:anomalous_samples}
\end{figure*}

\end{document}